\documentclass{aa}
\usepackage{float}
\usepackage{graphicx}
\usepackage{natbib}
\usepackage{txfonts}
\begin{document}

   \title{Smooth kinematic and metallicity gradients reveal that the Milky Way's nuclear star cluster and disc might be part of the same structure}

   
   

  \author{F. Nogueras-Lara
          \inst{1}       
           \and      
          A. Feldmeier-Krause
          \inst{2}         
           \and      
          R. Schödel
          \inst{3}         
           \and      
          M. C. Sormani
          \inst{4}             
           \and      
          A. de Lorenzo-Cáceres
          \inst{5,6}             
          \and
          A. Mastrobuono-Battisti
          \inst{7}  
            \and      
          M. Schultheis
          \inst{8}  
          \and
          N. Neumayer            
          \inst{2}      
          \and
          R. M. Rich 
           \inst{9}       
           \and
           N. Nieuwmunster       
          \inst{8}                                                            
          }

   \institute{
    European Southern Observatory, Karl-Schwarzschild-Strasse 2, D-85748 Garching bei M\"unchen, Germany
              \email{francisco.nogueraslara@eso.org}                         
     \and   
Max-Planck Institute for Astronomy, K\"onigstuhl 17, 69117 Heidelberg, Germany
\and       
     Instituto de Astrof\'isica de Andaluc\'ia (CSIC), Glorieta de la Astronom\'ia s/n, 18008 Granada, Spain            
     \and       
           Zentrum für Astronomie, Institut für theoretische Astrophysik, Universität Heidelberg, Albert-Ueberle-Str 2, D-69120 Heidelberg, Germany          \and
       Instituto de Astrofísica de Canarias, Calle Vía Láctea s/n, E-38205 La Laguna, Tenerife, Spain
       \and
       Departamento de Astrofísica, Universidad de La Laguna, E-38200 La Laguna, Tenerife, Spain
       \and
       GEPI, Observatoire de Paris, PSL Research University, CNRS, Place Jules Janssen, F-92190 Meudon, France
       \and
       Université Côte d'Azur, Observatoire de la Côte d'Azur, Laboratoire Lagrange, CNRS, Blvd de l'Observatoire, F-06304 Nice, France
       \and       
       Department of Physics and Astronomy, UCLA, 430 Portola Plaza, Box 951547, Los Angeles, CA 90095-1547
       }
   \date{Received 10 July 2023 / Accepted 25 October 2023}

 
  \abstract
   {The innermost regions of most galaxies are characterised by the presence of extremely dense nuclear star clusters. Nevertheless, these clusters are not the only stellar component present in galactic nuclei, where larger stellar structures known as nuclear stellar discs, have also been found. Understanding the relation between nuclear star clusters and nuclear stellar discs is challenging due to the large distance towards other galaxies which limits their analysis to integrated light. The Milky Way's centre, at only $\sim8$\,kpc, hosts a nuclear star cluster and a nuclear stellar disc, constituting a unique template to understand their relation and formation scenario.}
   {We aim to study the kinematics and stellar metallicity of stars from the Milky Way's nuclear star cluster and disc to shed light on the relation between these two Galactic centre components.} 
    {We used publicly available photometric, proper motions, and spectroscopic catalogues to analyse a region of $\sim2.8'\times4.9'$ centred on the Milky Way's nuclear star cluster. We built colour magnitude diagrams, and applied colour cuts to analyse the kinematic and metallicity distributions of Milky Way's nuclear star cluster and disc stars with different extinction, along the line of sight.}
   {We detect kinematic and metallicity gradients for the analysed stars along the line of sight towards the Milky Way's nuclear star cluster, suggesting a smooth transition between the nuclear stellar disc and cluster. We also find a bi-modal metallicity distribution for all the analysed colour bins, which is compatible with previous work on the bulk population of the nuclear stellar disc and cluster. Our results suggest that these two Galactic centre components might be part of the same structure with the Milky Way's nuclear stellar disc being the grown edge of the nuclear star cluster.}  
   {}

\titlerunning{Evidence of an age gradient along the line of sight in the NSD}
\authorrunning{F. Nogueras-Lara et al.}
   \maketitle
%

\section{Introduction}


The innermost regions of most galaxies host a nuclear star cluster (NSC). NSCs are extremely dense and massive stellar clusters that are normally identified as luminous compact sources different from their surrounding area \citep{Neumayer:2020aa}. These clusters are not the only stellar structure that can be present in galactic nuclei. In particular, the TIMER survey analysed the inner few kpc of 21 nearby massive and barred galaxies and found the presence of nuclear stellar discs (NSDs) in 19 of them \citep{Gadotti:2020aa}. These stellar structures are significantly larger than NSCs and their formation is attributed to gas that was funnelled by the galactic bar towards the innermost regions of the galaxies \citep{Bittner:2020aa}. NSCs and NSDs can co-exist in galactic nuclei \citep[e.g.][]{Lyubenova:2013aa}, although their relationship is not yet well studied.

The Milky Way's centre is the closest galaxy nucleus, located at only $\sim8$\,kpc from Earth \citep[e.g.][]{Gravity-Collaboration:2018aa,Do:2019aa}, and the only one where it is possible to resolve individual stars down to milli-parsec scales \citep[e.g.][]{Schodel:2010fk,Nogueras-Lara:2018aa}. It hosts the Milky Way's nuclear star cluster (MWNSC), which has a total stellar mass of $\sim 2.5 \times 10^7$\,$M_\odot$ \citep[e.g.][]{Launhardt:2002nx,Schodel:2014bn,Feldmeier:2014kx,Chatzopoulos:2015yu,Fritz:2016aa,Feldmeier-Krause:2017tk} and an effective radius of $\sim 4-5$\,pc \citep[e.g.][]{Schodel:2014fk,gallego-cano2019}. It is embedded in the Milky Way's nuclear stellar disc (MWNSD), a much larger and flatter stellar structure with a total mass of $\sim 10^9$\,$M_\odot$ \citep[e.g.][]{Launhardt:2002nx,Nogueras-Lara:2019ad,Sormani:2020aa,Sormani:2022wv}, a scale length of $\sim100$\,pc, and a scale height of $\sim40$\,pc \citep[e.g.][]{Launhardt:2002nx,gallego-cano2019,Sormani:2022wv}. Hence, an NSC and an NSD coexist in the Milky Way, which can thus serve as a laboratory to understand the relation between these Galactic centre components. 

In spite of their proximity, the study of the MWNSC and the MWNSD is hampered by the extreme source crowding and the high extinction that characterise the Galactic centre \citep[e.g.][]{Scoville:2003la,Nishiyama:2008qa,Fritz:2011fk,Chatzopoulos:2015uq}. Understanding the relation between the MWNSC and the MWNSD is challenging due to the 2D projection effects, and requires an analysis of their stellar populations along the line of sight towards the MWNSC. Hence, it is necessary to obtain clean samples of stars from each component to be able to determine their properties. 

Recent photometric studies have shown that the MWNSC and the MWNSD appear to contain different stellar populations and may have experienced independent formation histories \citep{Nogueras-Lara:2019ad,Schodel:2020aa,Nogueras-Lara:2021wm}. Although both structures seem to be dominated by an old stellar population, around 15\,\% of the total MWNSC stellar mass formed $\sim3$\,Gyr ago, while the central regions of the MWNSD underwent a quiescence phase with little star formation in between $\sim2-7$\,Gyr ago \citep{Nogueras-Lara:2023aa}. Moreover, around $5$\,\% of the total stellar mass of the MWNSD formed in a very energetic and short ($\sim100$\,Myr) star formation episode $\sim1$\,Gyr ago \citep{Nogueras-Lara:2019ad}, which did not happen in the MWNSC. On the other hand, a recent spectrophotometric analysis of the star formation history of the NSC suggests that the dominant stellar population could be younger that previously found \citep[$5\pm3$\,Gyr old,][]{Chen:2023aa}, implying an even starker difference in comparison to the MWNSD.

\citet{Nogueras-Lara:2022tp} found an extinction gradient along the line of sight towards the Galactic centre and used it to identify differences between mean kinematics and metallicity of the MWNSC and the MWNSD. The MWNSD was found to rotate faster than the MWNSC, and to have a lower metallicity than the MWNSC, which is arguably the most metal rich region of the Galaxy. 

Nevertheless, previous studies analysed the average properties of the MWNSC and MWNSD, mixing stars located at very different Galactic centre radii along the line of sight due to the 2D projection effect \citep[e.g.][]{Schultheis:2021wf}. Therefore, considering the transition region between the MWNSC and the MWNSD along the line of sight towards the MWNSC is fundamental to understand their nature and to properly constrain their relation and apparently different properties. In this paper, we study the line of sight towards the MWNSC and find a correlation between extinction and distance of the stars from the Galactic centre. In this way, we are able to characterise the kinematics and metallicity of stars at different Galactic centre radii. We find that the transition between the MWNSD and the MWNSC is smooth and we detect the presence of kinematic and metallicity gradients towards the supermassive black hole at the heart of the MWNSC.

\section{Data}

\subsection{Photometry}

\begin{figure}
                 
   \includegraphics[width=\linewidth]{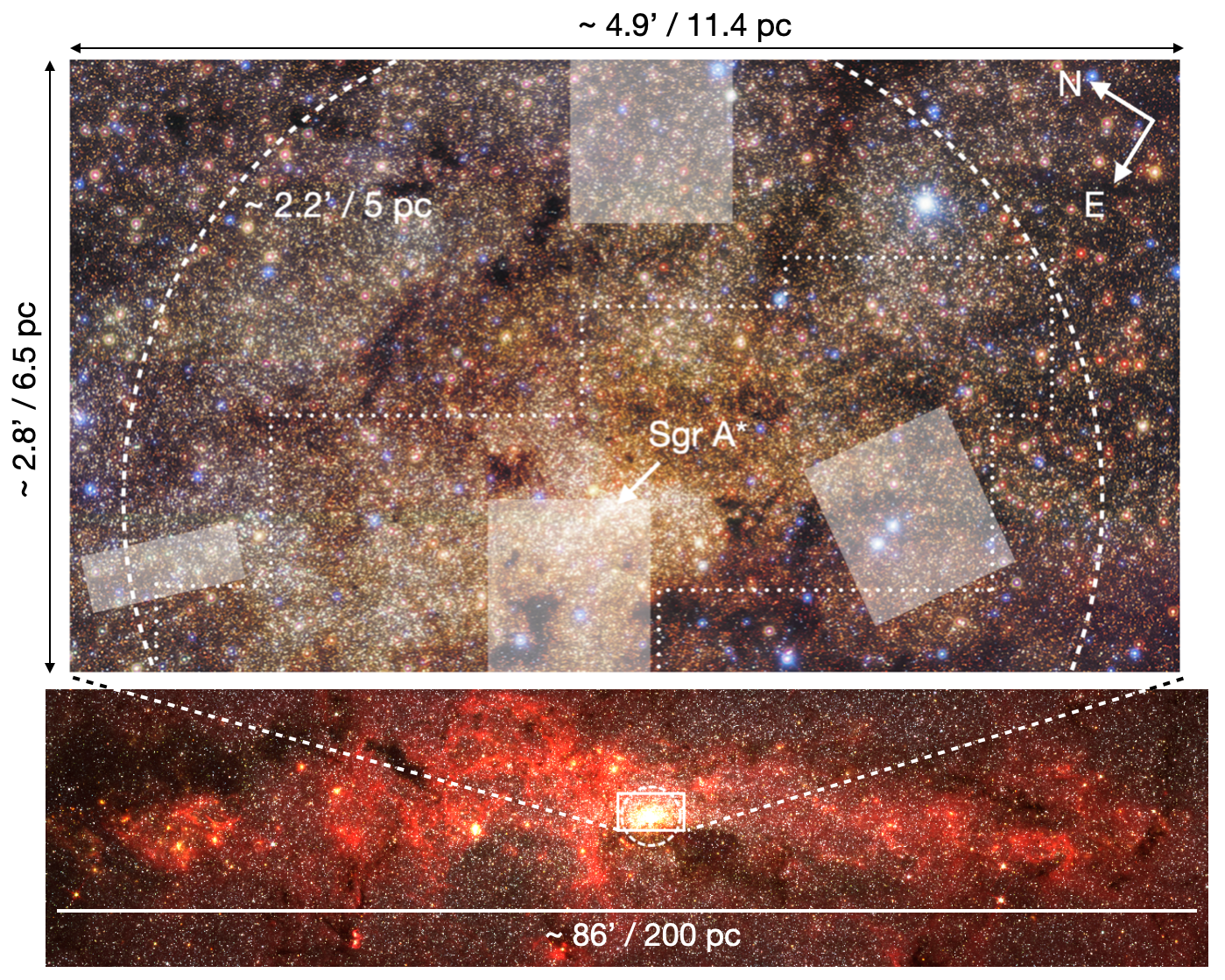}
   \caption{Target region centred on the MWNSC. The lower panel shows a Spitzer false colour image \citep[3.6\,$\mu$m (blue), 4.5\,$\mu$m (green), 5.8\,$\mu$m (red),][]{Stolovy:2006fk}. The upper panel corresponds to a false colour $JHK_s$ GALACTICNUCLEUS image adapted from \citet{Nogueras-Lara:2022tp}. The circular dashed region outlines the effective radius of the MWNSC, the white dotted contour shows the region with KMOS data, and the white filled boxes indicate regions without proper motion data. The position of the supermassive black hole Sagittarius\,A* (Sgr\,A*) and the approximate size of the fields at the Galactic centre distance are also indicated in the figure.}

\label{scheme1}
    \end{figure}

We used near infrared $JHK_s$ high-resolution ($\sim0.2''$) photometric observations of a region of $\sim2.8'\times4.9'$ ($\sim6.5$\,pc $\times$ 11.4\,pc at the Galactic centre distance) centred on the MWNSC, obtained with the HAWK-I camera at the VLT \citep{Kissler-Patig:2008uq}. Figure\,\ref{scheme1} shows the analysed region and its position with respect to the MWNSD. The observations consisted in a series of short-exposures frames (DIT = 1.26\,s for $J$, and DIT = 0.85\,s for $H$ and $K_s$) that were combined through applying the speckle holography algorithm \citep{Schodel:2013fk}. This reconstruction technique uses an averaged division of quantities in Fourier space (Eq.\,1 in \citealt{Schodel:2013fk}) to combine several hundreds of frames by convolving them with a Gaussian point spread function which defines the angular resolution of the final image (FWHM of $0.2''$). 

The $J$-band data correspond to the GALACTICNUCLEUS survey (Field 1 observed on 6 June 2015, see Table\,A.1 in \citet{Nogueras-Lara:2019aa} for further details), and the $H$- and $K_s$-band data correspond to a pilot study for the same survey \citep{Nogueras-Lara:2018aa} obtained under significantly better conditions than the actual survey (seeing in $K_s\sim0.4''$, data observed on 7 June 2013), which allowed us to achieve photometric measurements $\sim1$\,mag deeper than the GALACTICNUCLEUS catalogue. The $K_s$-band data presented significant saturation problems for $K_s<11.5$\,mag, which were corrected using the SIRIUS IRSF survey \citep{Nagayama:2003fk,Nishiyama:2006tx}, as explained in \citet{Nogueras-Lara:2019ad}.

\subsection{Proper motions}

We used the proper motion catalogue from \citet{Shahzamanian:2021wu} which covers the majority ($\sim80\,\%$) of the target region (Fig.\,\ref{scheme1}). The proper motions were obtained combining the $H$-band data from the GALACTICNUCLEUS survey \citep{Nogueras-Lara:2018aa,Nogueras-Lara:2019aa}, and the HST Paschen-$\alpha$ survey \citep{Wang:2010fk,Dong:2011ff}, with a timeline of 7-8 years between them. 

The computed proper motions are relative, meaning that they were calculated assuming an average zero motion of the stellar population. Hence, an offset exists between the proper motions with respect to a reference frame that is at rest at Sgr\,A*, and the ones in the catalogue, which depends on the real average motion of the stellar population \citep[mainly dominated by stars moving eastwards, e.g.][]{Trippe:2008it,Chatzopoulos:2015yu,Shahzamanian:2021wu,Martinez-Arranz:2022uf,Nogueras-Lara:2022tp}. To correct this offset affecting the proper motion component parallel to the Galactic plane ($\mu_l$), we applied a correction of $+0.36$\,mas/yr for the $\mu_l$ calculation when using the CMD $K_s$ versus $J-K_s$ to select stars belonging to the MWNSC and the MWNSD. This correction was computed by \citet{Nogueras-Lara:2022tp} by applying a Gaussian Mixture model \citep[GMM, ][]{Pedregosa:2011aa} to fit the $\mu_l$ distribution of the MWNSC stars. This distribution is best represented by a three-Gaussians model corresponding to stars moving eastwards and westwards, and also to a central, more slowly rotating component. They estimated the correction for the relative proper motion calculation as the offset of the central Gaussian component with respect to zero. Applying this correction to the mean values of the Gaussian corresponding to the stars moving eastwards and westwards, they obtained a similar absolute $\mu_l$ value for each component, as expected \citep[see Table\,1 and Sect.\,3.3 in][]{Nogueras-Lara:2022tp}.

We applied an analogous process to estimate the offset for the proper motion calculation when using the CMD $K_s$ versus $H-K_s$ to choose stars from the MWNSC and the MWNSD. We obtained an offset of $\mu_l=0.5$\,mas/yr. This value is slightly larger than the one obtained by \citet{Nogueras-Lara:2022tp} when using the CMD $K_s$ versus $J-K_s$ as a reference, probably due to the higher completeness in the $HK_s$ bands. In any case, the presence of a $\mu_l$ offset does not pose any problems for the subsequent analysis because we compare the MWNSC and the MWNSD using data along the same line of sight, and thus this potential offset will affect all stars in the field in a similar way.

\subsection{Metallicity and line-of-sight velocities}

We used the KMOS spectroscopic data obtained by \citet{Feldmeier-Krause:2017kq,Feldmeier-Krause:2020uv}. They analysed $K$-band medium-resolution spectra using a full spectral fitting method with PHOENIX models \citep{Husser:2013uu}.

They also obtained line-of-sight velocities using the STARKIT code developed by \citet{Kerzendorf:2015aa}. We cross-correlated our photometry with the spectroscopic data and ended up with $\sim1000$ stars with measured $[M/H]$ and line-of-sight velocities in the target field.

\section{Kinematics along the line of sight}
\label{crite}

\begin{figure}
   \includegraphics[width=\linewidth]{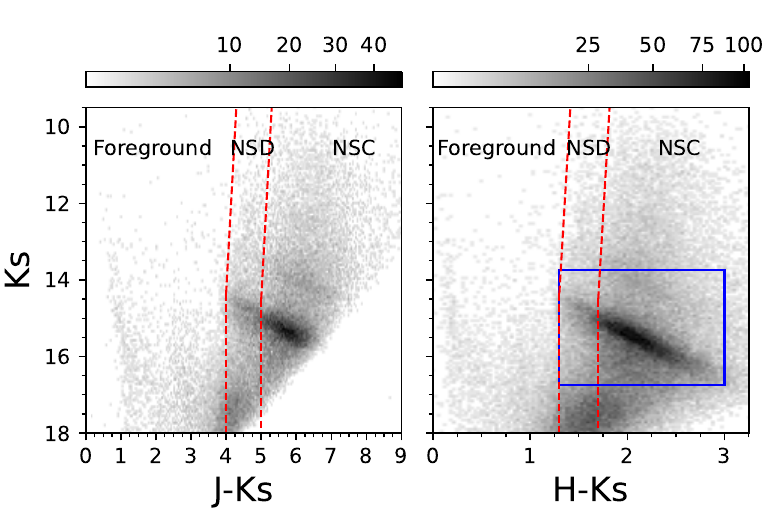}
   \caption{CMDs $K_s$ versus $J-K_s$ (left panel) and $K_s$ versus $H-K_s$ (right panel) for the target region. The red dashed lines indicate the separation between the foreground population, the MWNSD, and the MWNSC. The blue box in the CMD $K_s$ versus $H-K_s$ shows the reference stars to build the extinction map in Sect\,\ref{sect_map}. The scale bars indicate stellar densities using a power stretch scale.}

   \label{CMD}
    \end{figure}

Previous work analysing the line of sight towards the MWNSC determined that the MWNSD and the MWNSC can be distinguished via their different extinction \citep{Nogueras-Lara:2021wm,Nogueras-Lara:2022tp}. Figure\,\ref{CMD} shows the colour magnitude diagrams (CMDs) $K_s$ versus $J-K_s$ and $K_s$ versus $H-K_s$, where the separation between the MWNSD and the MWNSC is indicated in agreement with previous studies. This allows us to distinguish between these two Galactic centre components in the CMDs. Moreover, the extreme extinction towards the Galactic centre also helps us exclude foreground stars which mainly belong to the disc of the Milky Way, and to some extent to the Galactic bar \citep[e.g.][]{Nishiyama:2008qa,Gonzalez:2012aa,Surot:2020vo,Nogueras-Lara:2021uz}. 

Identifying the MWNSD and the MWNSC along the line of sight towards the MWNSC implies a statistical correlation between extinction and distance. The stars from the near side of the MWNSD are on average less extinguished than stars from the MWNSC along the same line of sight, and can be identified by their different colour. In this work, we aim to go further and analyse the kinematic behaviour of the stellar population from the MWNSD and the MWNSC by applying colour cuts to evenly sample the colour space. In this way, we consider MWNSD and MWNSC stars located at different Galactic centre radii instead of averaging over all the stars from each of the components, as done in previous work \citep[e.g.][]{Schultheis:2021wf,Nogueras-Lara:2021wm,Nogueras-Lara:2022tp}.

\subsection{CMDs and proper motion distribution}
\label{proper_sect}

We cross-correlated the photometric and proper motion catalogues for the target region, and restricted our analysis to stars with proper motion uncertainties $<0.5$\,mas/yr to avoid potential bias due to stars with large kinematic uncertainties. We ended up with $\sim2100$, and $\sim2500$ stars with well defined proper motions and a photometric counterpart in $JK_s$ and $HK_s$, respectively. The lower number of stars with photometric counterpart in $JK_s$ is because shorter wavelengths are more affected by extinction and thus the $J$ catalogue is less complete. 

\begin{figure}
   \includegraphics[width=\linewidth]{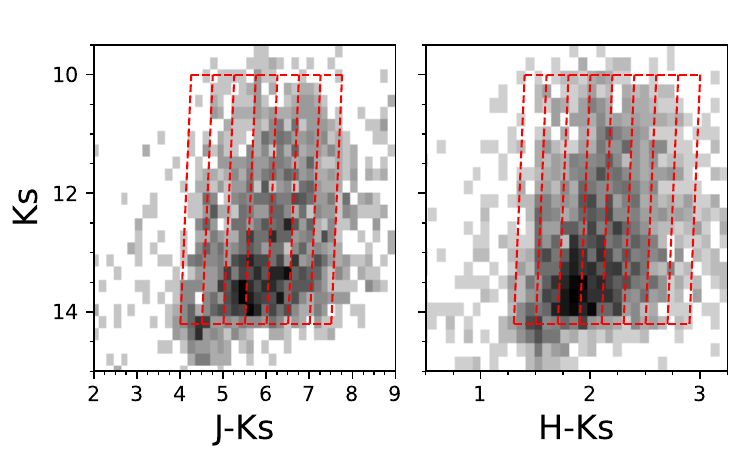}
   \caption{CMDs $K_s$ versus $J-K_s$ and $K_s$ versus $H-K_s$ obtained for stars with well defined proper motions (i.e. proper motion uncertainties $<0.5$\,mas/yr). The red dashed boxes indicate the assumed colour cuts to analyse the variation of the proper motion distribution with respect to the extinction.}

   \label{CMD_pm}
    \end{figure}

Figure\,\ref{CMD_pm} shows the CMDs $K_s$ versus $J-K_s$ and $K_s$ versus $H-K_s$ for the common stars between the used catalogues. To analyse the proper motion distribution, we defined colour bins for each CMD with a width equivalent to $A_{K_s}\sim0.25$\,mag. This corresponds to a variation in colour of $J-K_s=0.5$\,mag and $H-K_s=0.2$\,mag assuming the extinction curve $A_J/A_{K_s}=1.84$ and $A_J/A_{K_s}=3.44$  \citep{Nogueras-Lara:2020aa}. The inclination of the colour bins was chosen to be parallel to a stellar isochrone of $\sim10$\,Gyr which corresponds to the dominant old stellar population in both the MWNSD and the MWNSC \citep[e.g.][]{Nogueras-Lara:2019ad,Schodel:2020aa}, in agreement with previous work \citep[see Fig.\,3 in][]{Nogueras-Lara:2022tp}.

We considered the data completeness for the bright and the faint end of the colour cuts, and limited our analysis to stars with $K_s\in[10,14.2]$\,mag \citep{Nogueras-Lara:2022tp}. Given the higher completeness for the CMD $K_s$ versus $H-K_s$, we selected 8 colour bins instead of the 7 chosen for the CMD $K_s$ versus $J-K_s$.

\subsubsection{MWNSD and MWNSC in the CMDs}

We checked the consistency of our statistical distinction between MWNSD and MWNSC stars when applying the colour cuts $J-K_s$ and $H-K_s$, as specified in Fig.\,\ref{CMD}. For this, we used all the stars fulfilling the  previously specified criteria and with available $JHK_s$ photometry. We obtained that the agreement between stars identified as belonging to the MWNSD in the CMDs $K_s$ versus $J-K_s$ and $K_s$ versus $H-K_s$ is $\gtrsim 85$\,\%, whereas this number rises up to $\gtrsim 95$\,\% in the case of MWNSC stars. Therefore, we conclude that our selection criterion is consistent regardless of the colour used.

\subsubsection{Results}
        
\begin{figure}
   \includegraphics[width=\linewidth]{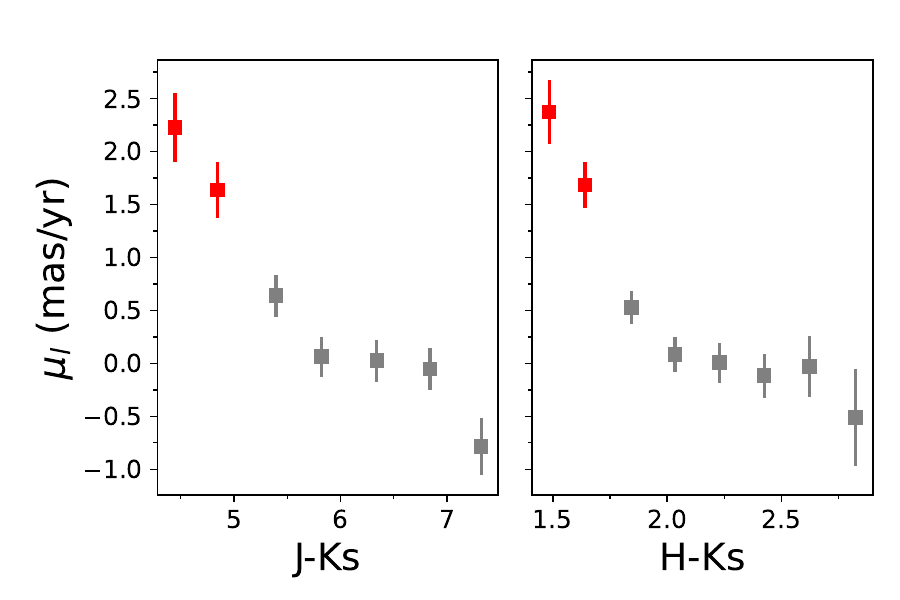}
   \caption{Variation of the proper motion component parallel to the Galactic plane versus colour. The proper motion values were calculated assuming the median value of the stars in each colour bin. The red and grey squares correspond to stars from the MWNSD and MWNSC following the extinction criterion in Sect.\,\ref{crite}.}

   \label{gradient_proper_motions}
    \end{figure}

We computed the median value of the proper motion components parallel and perpendicular to the Galactic plane for each of the colour bins previously defined. Tables\,\ref{proper_motions_JK} and \ref{proper_motions_HK} show the results. The uncertainties, $d\mu_i$, were estimated assuming the standard error of the median values, where the subindex $i$ indicates proper motion component parallel ($l$), or perpendicular ($b$) to the Galactic plane. Figure\,\ref{gradient_proper_motions} shows the behaviour of the median $\mu_l$ value for the selected colour cuts. We found a clear dependence of $\mu_l$ with respect to the applied colour cuts, which agrees for both CMDs.


Our results are compatible with the eastwards rotation of the Galactic centre stellar components \citep[e.g.][]{Trippe:2008it,Chatzopoulos:2015yu,Shahzamanian:2021wu,Sormani:2022wv}. Following previous work \citep{Nogueras-Lara:2021wm,Nogueras-Lara:2022tp}, we distinguished between the stars likely belonging to the MWNSD and the MWNSC by applying an extinction criterion, as indicated in Fig.\,\ref{CMD}. In this way, the red and grey squares in Fig.\,\ref{gradient_proper_motions} show the $\mu_l$ distribution for MWNSD and MWNSC stars, respectively. We observed that the $\mu_l$ velocity of the MWNSD is maximum for stars from its outer edge \citep{Nogueras-Lara:2022aa,Nogueras-Lara:2023aa}, whereas stars from its innermost regions present lower $\mu_l$ values. This tendency continues for MWNSC stars, and reaches a zero value at $J-K_s\sim6.5$\,mag, and $H-K_s\sim2.5$\,mag, probably dominated by stars close to Sgr\,A*, which are mainly pressure-supported \citep[see ][]{Trippe:2008it,Schodel:2009zr}. This $\mu_l$ behaviour is also in agreement with the minimum value obtained for the line-of-sight velocity of MWNSC stars close to Sgr\,A* \citep[see Fig.\,6 in ][]{Nogueras-Lara:2022tp}.

The detected variation of $\mu_l$ with colour (i.e. extinction) also validates our initial assumption of the statistical identification of stars from different Galactic centre radii along the line of sight towards the MWNSC by applying colour cuts in the CMDs. Our analysis does not imply that each assumed colour cut is a clean sample of stars at different Galactic centre radii. It is a statistical approach meaning that the majority of stars in the sample are at larger distances along the line of sight for higher extinctions.

Previous work analysed the bulk $\mu_l$ distribution of the MWNSD and the MWNSC and found that, in both cases, they are best represented by the combination of three Gaussians \citep{Shahzamanian:2021wu,Martinez-Arranz:2022uf,Nogueras-Lara:2022tp}. They obtained a rotation velocity of $\sim2.3$\,mas/yr and $\sim 2.1$\,mas/yr \citep[after correcting for the relative proper motion calculation,][]{Nogueras-Lara:2022tp}, for the MWNSD and the MWNSC, respectively. An average of our results for the corresponding stellar populations is somewhat smaller than previous results. This is explained by the different methodology (we computed median values) and the potential contamination from stars deeper inside the line of sight, that is accounted by a GMM approach in previous work. Our analysis is also able to detect the differential rotation of the MWNSC \citep[e.g.][]{Trippe:2008it,Schodel:2009zr}, which is shown by the different $\mu_l$ values obtained at different colour bins  in Fig.\,\ref{gradient_proper_motions}.

We also analysed the median values of the proper motion component perpendicular to the Galactic plane ($\mu_b$). We found that they do not depend on colour and are close to zero (Tables\,\ref{proper_motions_JK} and \ref{proper_motions_HK}). Our results are compatible with previous work \citep[e.g.][]{Shahzamanian:2021wu, Nogueras-Lara:2022aa}.

\subsubsection{Completeness analysis}
\label{comp}

To check the impact of the completeness on the detected $\mu_l$ gradient, we assumed that our $K_s$ photometric catalogue is fully complete for stars with $K_s\in[10,14.2]$\,mag in the analysed region \citep{Nogueras-Lara:2018aa}. We built a $K_s$-luminosity function with the previously used stars with well defined proper motions (uncertainty below 0.5\,mas/yr), and compared it with an analogous luminosity function created using all the stars in our photometric catalogue (see cut in Fig.\,\ref{CMD}). Given the different spatial coverage of the catalogues, we excluded regions which were not covered by our proper motion catalogue (white filled boxes in Fig.\,\ref{scheme1}) to create the photometric-data luminosity function. We obtained that the proper motion completeness is $\gtrsim60$\,\% in the range $K_s\in[10,13]$\,mag. Finally, we repeated our $\mu_l$ analysis considering only stars in that brightness range to guarantee a large completeness. We obtained similar results within the uncertainties and concluded that the completeness does not impact our results in any significant way.

\subsection{Velocity dispersion}
\label{bootstrap}
\begin{figure}
   \includegraphics[width=\linewidth]{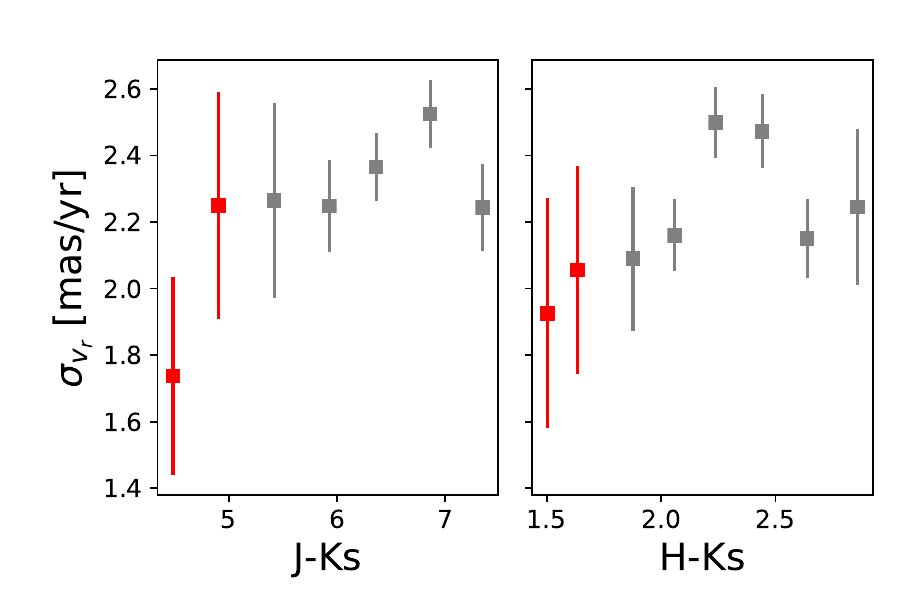}
   \caption{Variation of the dispersion of line-of-sight velocities as a function of colour. The red and grey squares correspond to stars from the MWNSD and MWNSC following the extinction criterion in Sect.\,\ref{crite}. The $x$-coordinates of the data points were calculated assuming the median value of the stars in each colour bins following the extinction criterion in Sect.\,\ref{crite}.}

   \label{gradient_vr}
    \end{figure}

We complemented our previous analysis by computing the distribution of the velocity dispersion for the $\sim1000$ KMOS stars  with available photometry and line-of-sight velocities. We used this stellar sample because line-of-sight velocities were obtained with a considerably lower uncertainty \citep[$\lesssim 0.2$\,mas/yr, ][]{Feldmeier-Krause:2017kq,Feldmeier-Krause:2020uv} in comparison to the proper motion uncertainties for each star (<0.5\,mas/yr, see Sect.\,\ref{proper_sect}). We applied the same colour cuts indicated in Fig.\,\ref{CMD_pm} and analysed the velocity dispersion of the KMOS stars. We estimated the velocity dispersion of the stars within each colour bin and its corresponding uncertainty by applying a bootstrap resampling method with 1000 iterations. We created 1000 samples by randomly resampling the original data set with replacement. We estimated the velocity dispersion and its associated uncertainty as the mean and the standard deviation of the result computed for each sample. Table\,\ref{table_vr} and Fig.\,\ref{gradient_vr} show the obtained results.

We observe a smooth kinematic variation between the MWNSD and the MWNSC, which shows an increase of the velocity dispersion with respect to the colour (i.e. extinction), reaching a maximum for $J-K_s\sim7$\,mag and $H-K_s\sim2.3$\,mag and then decreasing. This is compatible with the correlation between the extinction and the position of the stars along the line of sight, and agrees with the larger velocity dispersion expected for stars close to Sgr\,A* \citep[e.g.][]{Trippe:2008it,Schodel:2009zr,Feldmeier:2014kx}.

To assess our results and check the maximum velocity dispersion values that we obtained, we repeated our analysis considering only stars in a projected radius of $\sim1.25'$ ($\sim3$\,pc at the Galactic centre distance) from Sgr\,A*. In this way, we reduced the contamination from stars with similar extinction to the ones near Sgr\,A*, but located at a much larger distance due to the 2D projection. We did not observe any significant difference within the uncertainties.

Our results are compatible within the uncertainties with the mean velocity dispersion profile for the MWNSC and the inner regions of the MWNSD obtained by \citealt{Feldmeier-Krause:2022vm} (see their Fig.\,12). Moreover, the maximum velocity dispersion value also agrees with previous values for the innermost MWNSC $\sim10''-40''$, equivalent to $\sim0.4-1.5$\,pc at the Galactic centre distance \citep[see Fig.\,7 in ][]{Fritz:2016aa}.

The larger uncertainties obtained for the two first data points corresponding to MWNSD stars might be due to the lower number of stars in these colour bins and/or some residual contamination from the Galactic bulge/bar. This contamination could account for up to 20\,\% of the MWNSD stars with $H-K_s>1.3$\,mag \citep[see Table\,2 in][]{Sormani:2022wv}. Due to the expected extinction for Galactic bulge/bar stars \citep[e.g.][]{Gonzalez:2012aa,Nogueras-Lara:2018ab,Surot:2020vo}, and the impossibility to observe stars from the Galactic bulge/bar beyond the MWNSC because of the extreme extinction \citep[e.g.][]{Schodel:2010fk,Chatzopoulos:2015uq}, this contamination mainly affects the low extinction regime of the analysed field (i.e. MWNSD stars).

To check our results, we repeated the same analysis computing the velocity dispersion distribution of the proper motion components $\mu_l$ and $\mu_b$. We obtained consistent results although the uncertainties are larger than the ones obtained when considering line-of-sight velocities, as it was expected given the larger proper motion errors ($<0.5$\,mas/yr).

               \begin{figure*}
   \includegraphics[width=\linewidth]{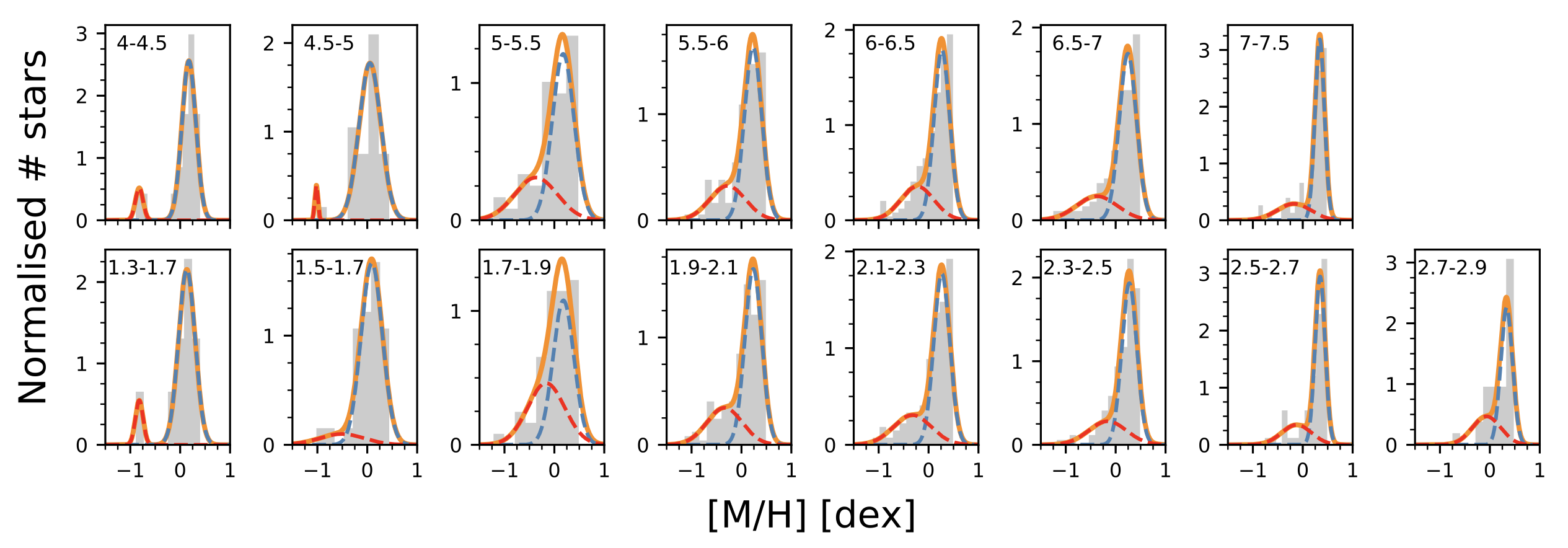}
   \caption{Metallicity distribution for the stars belonging to each colour bin in $J-K_s$ (upper row), and $H-K_s$ (lower cut). As we move from left to right, we move deeper inside the MWNSD along the line of sight. The range of the colour bins is indicated in each panel. The orange line shows the obtained GMM model for each metallicity distribution, whereas the red and blue dashed lines correspond to each of the Gaussian models.}

   \label{met_grad}
    \end{figure*}

\section{Metallicity}

We used the $\sim1000$ KMOS stars \citep{Feldmeier-Krause:2017kq,Feldmeier-Krause:2020uv} with available metallicity and photometry. Given that the spectral fitting was calibrated using existing empirical spectra for $[M/H] < 0.3$\,dex, the metallicities for stars with $[M/H] > 0.3$\,dex could have been overestimated. To asses the results, \citet{Schultheis:2021wf} applied the metallicity-spectral index relation derived by \citet{Fritz:2020aa} on empirical spectra of stars with $[M/H] < 0.6$\,dex. They found consistent result with the full spectral fitting method using synthetic spectra for $[M/H] \lesssim 0.5$\,dex.
In addition, \citet{Feldmeier-Krause:2022vm} tested the full spectral fitting technique on empirical spectra over a wide range of metallicities, and found consistent results with the literature at $[M/H] < 0.5$\,dex. Therefore, for this study we restricted our subsequent study to stars with $[M/H]<0.5$\,dex, as done in Sect.\,5 of \citet{Nogueras-Lara:2022tp}. We applied the same colour cuts previously specified in Fig.\,\ref{CMD_pm} and carried out an analysis of the metallicity distribution for each colour bin.

\subsection{Metallicity distribution}
\label{met_analysis}

Previous work indicates the presence of a bi-modal metallicity distribution for both, the MWNSD and the MWNSC \citep{Schultheis:2021wf,Nogueras-Lara:2022tp}. Thus, we used a GMM \citep{Pedregosa:2011aa} to derive the number of Gaussian models which best reproduces the metallicity distribution in each colour bin $J-K_s$ and $H-K_s$. We distinguished between one and two Gaussians by applying the Akaike information criterion \citep[AIC, ][]{Akaike:1974aa}, and obtained that for all the colour bins in $J-K_s$ and $H-K_s$ the probability density function of the data is always best represented by a two-Gaussians model. Figure\,\ref{met_grad} shows the results. To estimate the final GMM values and their associated uncertainties, we resorted to MonteCarlo (MC) simulations and repeated the GMM approach assuming a bi-modal distribution on 1000 MC samples generated randomly varying the metallicity of each star, assuming Gaussian uncertainties. Tables\,\ref{met_table_JK} and \ref{met_table_HK} show the results, where the mean values and their uncertainties were obtained averaging over the 1000 MC samples and applying a three sigma outlier-resistant criterion.

\begin{figure}
   \includegraphics[width=\linewidth]{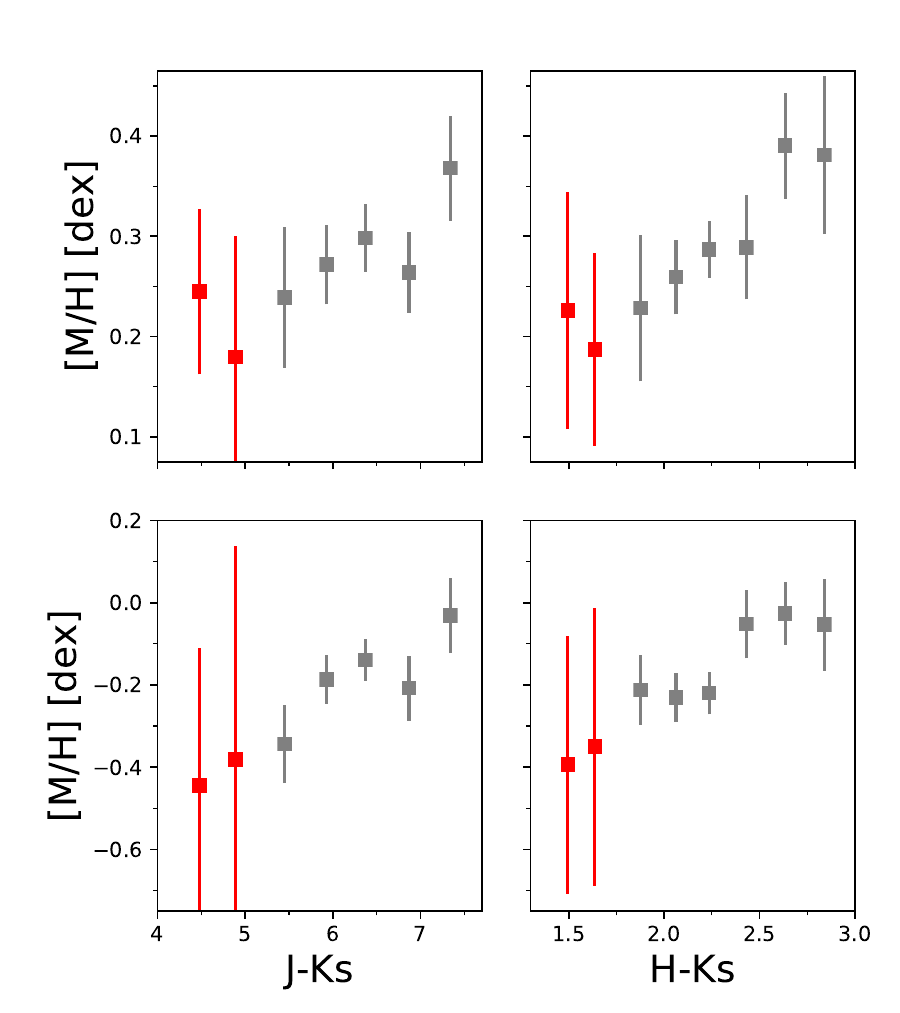}
   \caption{Metallicity as a function of colour. Upper panels: Metal rich stellar population. Lower panels: Metal poor stellar population. The red and grey squares correspond to stars from the MWNSD and MWNSC following the extinction criterion in Sect.\,\ref{crite}.}
   
   \label{gradient_met}
    \end{figure}

Figure\,\ref{gradient_met} shows the variation of the metallicity for the metal poor and rich components with respect to the applied colour cuts. We obtained that both stellar populations follow an increasing trend towards high colour values whose maximum is approximately at $J-K_s\sim6.5-7$ and $H-K_s\sim2.5$\,mag, which is in agreement within one sigma uncertainties with the position of Sgr\,A* that we derived in our kinematic analysis (i.e. $\mu_l\sim0$\,mas/yr, see Sect.\,\ref{crite}).

The most metal rich values obtained for MWNSC stars with high extinction, are somewhat larger than the metal rich peak obtained in previous work using the same data set \citep{Nogueras-Lara:2022tp} but analysing the mean properties of the MWNSC without considering colour cuts. This is consistent with the presence of a metallicity gradient, and can be justified by the averaged values obtained in previous work combining stars at different MWNSC radii.

\subsection{Completeness analysis}
\label{comp_met}

We estimated the completeness of our KMOS sample following the technique described in Sect.\,\ref{comp}. We created $K_s$ luminosity functions for the KMOS stars and for all the stars with $K_s$ photometry in the same region excluding foreground stars (see Fig.\,\ref{scheme1}). Assuming that our $K_s$ photometry is fully complete for the magnitude range covered by the KMOS data, we obtained that the completeness of the KMOS sample is $\sim50$\,\% for $K_s\in[10,11.5]$\,mag. The slightly lower completeness obtained in comparison with the one estimated by \citet{Feldmeier-Krause:2020uv} is because we removed stars with $[M/H]>0.5$\,dex, as it was previously explained.

To check the effect of completeness on the obtained metallicity gradients, we repeated the analysis considering only stars with $K_s\in[10,11.5]$\,mag. Given the lower number of stars in this magnitude bin, we doubled the size of the colour bins to have a sufficient number of stars per colour bin. We obtained similar results within the uncertainties and thus concluded that the completeness is not affecting our results.


\subsection{Spatial distribution of stars with different metallicities}

\begin{figure}
   \includegraphics[width=\linewidth]{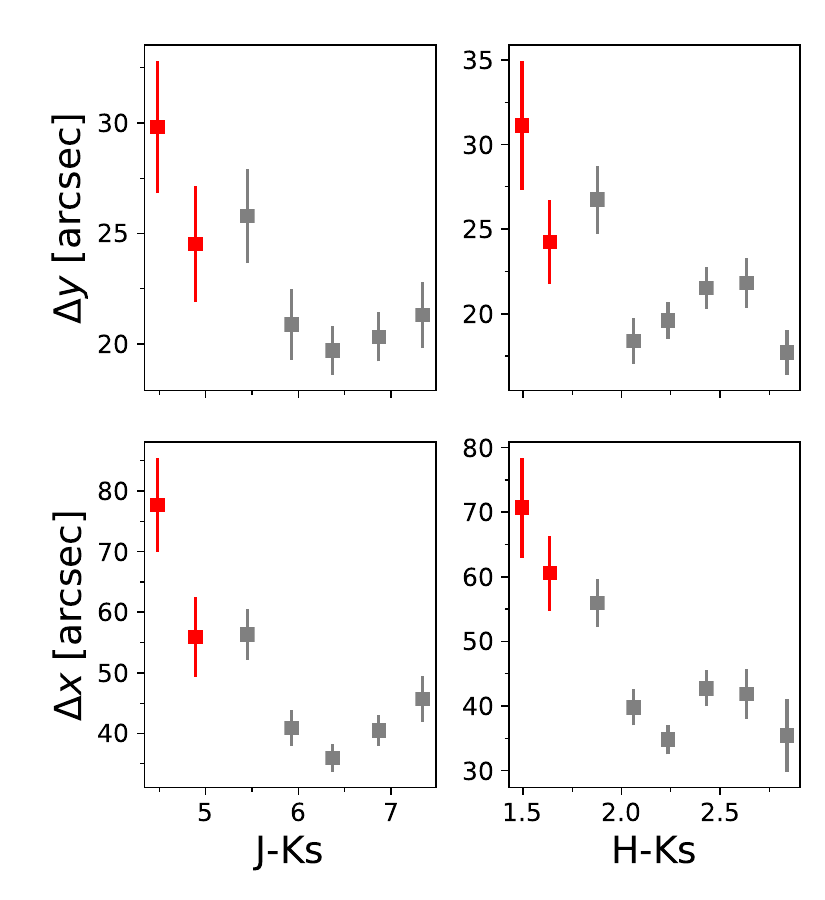}
   \caption{Standard deviation of the stellar positions of KMOS stars for each colour bin in $J-K_s$ and $H-K_s$. The red and grey squares correspond to stars from the MWNSD and MWNSC following the extinction criterion in Sect.\,\ref{crite}.}

   \label{spatial}
    \end{figure}

To check whether the detected metallicity gradient along the line of sight towards Sgr\,A* is real, we also studied the spatial distribution of the KMOS stars depending on the defined colour bins. We computed the median and the standard deviation of the position of the stars in each colour bin in the CMDs $K_s$ versus $J-K_s$ and $K_s$ versus $H-K_s$, by applying a bootstrap resampling method as explained in Sect.\,\ref{bootstrap}. Figure\,\ref{spatial} and Table\,\ref{table_spatial} show the results. We obtained that the standard deviation of the position distribution in $x$ and $y$ ($\Delta x$ and $\Delta y$, respectively) tends to decrease for stars with higher extinction. Therefore these stars are more centrally concentrated than stars with lower extinction which potentially belong to the MWNSD and the external regions of the MWNSC. Our results are compatible with the presence of a correlation between extinction and different Galactic centre radii along the line of sight, and support that stars with $J-K_s\sim6.5$\,mag and $H-K_s\sim2.5-3$\,mag are probably associated to the innermost regions of the MWNSC. 

On the other hand, the average distribution of the stars is roughly centred on Sgr\,A*, although some scatter is detected which can be associated to data incompleteness (see Sect.\,\ref{comp_met}), differential extinction, and/or asymmetries in the stellar distribution for different colour cuts. This might be related to the presence of stars from recent merger events, as the one proposed in \citet{Feldmeier-Krause:2020uv} and \citet{Do:2020wr}.

\section{Discussion}


\subsection{Effect of differential extinction on our results}
\label{sect_map}

The extinction towards the Galactic centre is extreme and significantly varies along the line of sight \citep[e.g.][]{Nishiyama:2006tx,Nishiyama:2008qa,Schodel:2010fk,Chatzopoulos:2015uq,Nogueras-Lara:2018aa}. Hence, the differential extinction might affect the correlation that we found between distance and extinction, and also the detected kinematics and metallicity gradients. To assess our results, we built an extinction map to repeat our analysis excluding regions whose extinction is more than one standard deviation away from the median $A_{K_s}$ value of the extinction map.

                 \begin{figure*}
   \includegraphics[width=\linewidth]{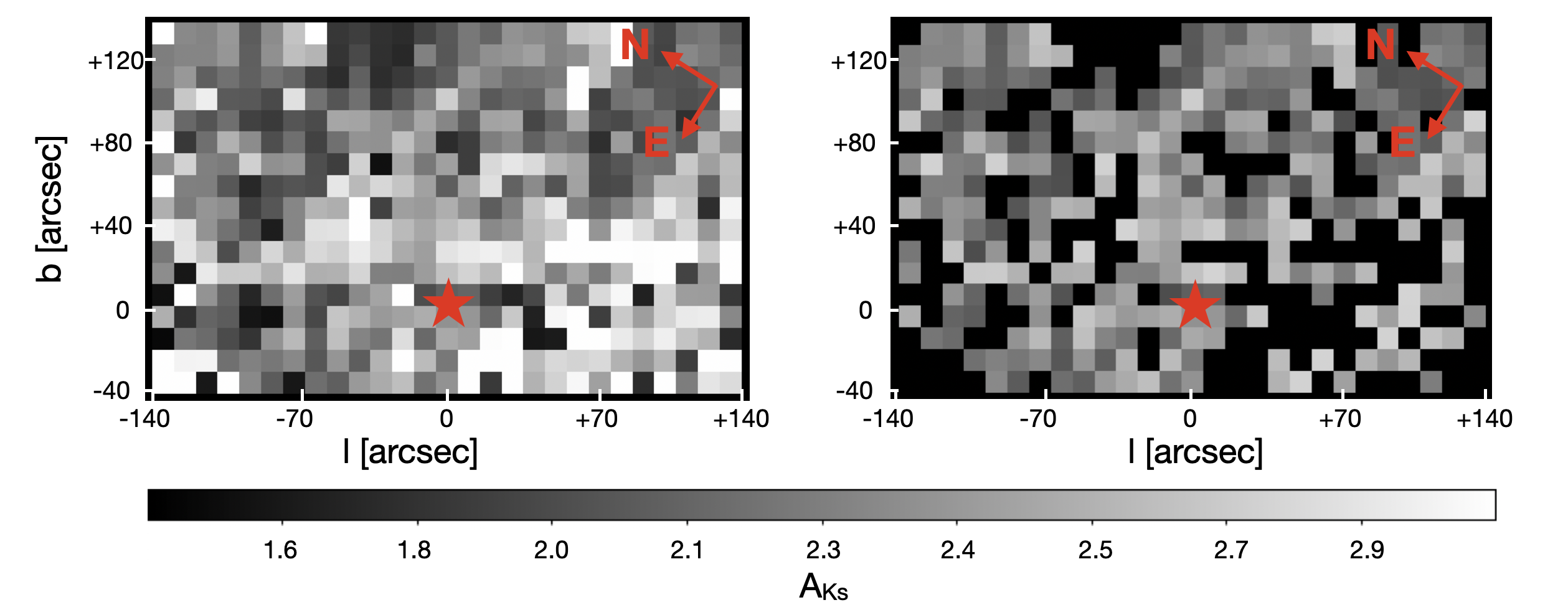}
   \caption{Extinction map for the entire analysed (left panel) and the same map excluding pixels (in black) that deviate by more than one standard deviation from the median extinction (right panel). The $x$ and $y$ axes indicate the distance from Sgr\,A* (red star) in arc-seconds. The extinction value of each pixel, $A_{K_s}$, is indicated by the scale bar.}

   \label{ext}
    \end{figure*}

We applied the technique described in \citet{Nogueras-Lara:2021wj} and used red clump stars \citep[helium core-burning stars with well characterised intrinsic properties, e.g.][]{Girardi:2016fk} and other red giant stars with similar intrinsic colours, as  reference stars. Figure\,\ref{CMD} shows the box that we used to choose the reference stars in the CMD $K_s$ versus $H-K_s$. We defined a pixel size of $\sim10''$ for the map and used the five closest reference stars to a given pixel within a radius of $\sim15''$ to compute its extinction using the equation:

\begin{equation}
A_{K_s} = \frac{H-K_s-(H-K_s)_0}{A_{H}/A_{K_s}-1},
\end{equation}

\noindent where $(H-K_s)_0 = 0.10\pm0.01$ is the intrinsic colour of the reference stars \citep[e.g.][]{Nogueras-Lara:2021wj}, and $A_{H}/A_{K_s}=1.84\pm0.03$ \citep{Nogueras-Lara:2020aa} is the extinction curve in the near infrared between the $H$ and $K_s$ bands. To account for the different distance of the reference stars from the centre of each pixel, we applied an inverse weight distance method \citep[for further details see ][]{Nogueras-Lara:2021wj}. To avoid mixing reference stars with too different extinctions, we constrained the maximum colour variation between the reference stars to be within 0.3\,mag from the central-most reference star for each pixel. We only assigned an extinction value to a given pixel if at least 5 reference stars were found.

We obtained a median extinction value of $A_{K_s} = 2.35$\,mag, with a standard deviation of $0.40$\,mag. Figure\,\ref{ext} shows the resulting extinction map (left panel), and the map after excluding the pixels more than one standard deviation away from the median extinction value (right panel). We repeated our $\mu_l$ analysis (see Sect.\,\ref{proper_sect}) excluding all the regions where the differential extinction varies more than one standard deviation, in agreement with the obtained map. We did not observe any variation within the uncertainties. Analogously, we also repeated the analysis of the metallicity gradient in Sect.\,\ref{met_analysis} and obtained similar results within the uncertainties. Therefore, we conclude that our analysis is robust and is not affected by the differential extinction in the analysed region.

An alternative way of assessing the effect of the differential extinction on our results is to increase the size of the colour bins in our analysis (see cuts in Fig.\,\ref{CMD_pm}). In this way, the colour bins contain stars with a wider range of extinction. We repeated our analysis of $\mu_l$, the velocity dispersion, and the metallicity gradients, doubling the size of the colour bins ($J-K_s=1$\,mag and $H-K_s=0.4$\,mag), and did not observe any significant difference in the results. Moreover, this approach allowed us to increase the number of stars per colour bin minimising potential problems in colour bins that previously had a relatively low number of stars.

\subsection{Stellar contamination between components}

The detected gradients suggest a smooth transition between the kinematic and metallicity properties of the MWNSC and the MWNSD, in contrast to previous analysis on their bulk population \citep[e.g.][]{Schultheis:2021wf,Nogueras-Lara:2022tp}. Nevertheless, this observation could be explained by a changing ratio of stars from these two Galactic centre components. To check this hypothesis, we focused on the metal rich stellar population and assumed that the MWNSD and the MWNSC are different components with different metallicities \citep[e.g.][]{Schultheis:2021wf,Nogueras-Lara:2022tp} equal to $[M/H]_{\mathrm{MWNSD}}\sim0.2$\,dex and $[M/H]_{\mathrm{MWNSC}}\sim0.4$\,dex (i.e. the extreme computed values in Tables\,\ref{met_table_JK} and \ref{met_table_HK}). 

We used all the stars detected in the photometric catalogue to avoid completeness issues and also considered the effect of the reddening vector to avoid including more stars in the bluer colour bins in comparison to the redder ones given the magnitude cut $K_s\in[10,14.2]$\,mag. We assumed that the two first colour bins in the CMDs in Fig.\,\ref{CMD_pm} have negligible or very low contamination from the MWNSC \citep[see Fig.\,9 in ][]{Sormani:2022wv}, and a maximum contamination of $\sim20$\,\% from Galactic bulge/bar stars \citep[e.g.][]{Sormani:2022wv}. For the remaining colour bins, we computed the number of stars from the MWNSD and the MWNSC that are necessary to obtain the metallicities in Tables\,\ref{met_table_JK} and \ref{met_table_HK}. We obtained that for both CMDs ($K_s$ versus $J-K_s$ and $K_s$ versus $H-K_s$), $\sim1800$ stars from the MWNSD are necessary to explain the detected gradient given the total number of $\sim3300$ stars in the samples. This means that $55$\,\% of the stars must belong to the MWNSD to justify the gradients.

To check whether this is possible, we estimated the 2D-projected mass density for the MWNSD and the MWNSC. For the MWNSC, we considered a distance of 3, 4, and 5\,pc from its centre and assumed the mass distribution presented in Fig.\,7 in \citet{Feldmeier-Krause:2017tk}. We obtained an average 2D-projected mass density of $\sim3\times 10^5$\,$M_\odot/$pc$^2$. For the MWNSD, we considered the total stellar mass of $\sim1.3\times10^8\,M_\odot$ obtained for a region of $\sim800$\,pc$^2$ near the MWNSC but excluding the region dominated by its effective radius \citep{Nogueras-Lara:2023aa}, and estimated a total MWNSD 2D-projected mass of $\sim1.6\times10^5\,M_\odot/$pc$^2$. To compute the final MWNSD 2D-projected mass density, we removed half of the mass because the extreme source crowding and extinction along the target line of sight impedes to observe the MWNSD beyond the MWNSC \citep[e.g.][]{Schodel:2010fk,Chatzopoulos:2015uq,Nogueras-Lara:2021wm}. We concluded that the stellar mass ratio between the MWNSD and the MWNSC is $\sim25$\,\% for the observed region, which is also in agreement with previous estimates \citep[e.g.][]{Nogueras-Lara:2021wm}. Therefore, although some mixing of stars from the MWNSD and MWNSC is expected, the contribution of the MWNSD stars in the region would need to be approximately double to fully account for the detected metallicity gradient. The same conclusion is also valid for the metal poor population, in which the variation between the extreme metallicity values is even larger ($\sim0.4$\,dex), and thus the mixture of different average populations from the MWNSD and the MWNSC is even more unlikely to explain the detected gradient.

To compare the estimated 2D-projected mass density and the number of stars from each component that are necessary to account for the detected gradients, we assumed that the MWNSD and MWNSC are both dominated by an old stellar population \citep{Nogueras-Lara:2019ad,Schodel:2020aa}, and thus differences in their stellar population will not cause any effect on the estimated stellar mass ratio of $\sim25$\,\% between the MWNSD and the MWNSC in the analysed magnitude range of $K_s\in[10,14.2]$\,mag.

The same analysis also applies to the potential effect of radial migration to mix the stellar populations. Although some mixture is expected, the number of stars is not enough to produce the observed metallicity gradient.

\subsection{Formation of the MWNSD and MWNSC}

We obtained a bi-modal metallicity distribution for all the colour cuts and bands used in our analysis. This indicates that the previously found bi-modality on the bulk population of the MWNSD and MWNSC \citep{Schultheis:2021wf,Nogueras-Lara:2022tp} is probably not a consequence of mixing of stellar populations located at different line-of-sight distance and/or from different Galactic centre components, but it is actually a real feature present in both the MWNSD and the MWNSC at different distances from Sgr\,A*. Moreover, the ratio between the metal poor and rich stars is approximately constant ($\sim0.6$, see $W_1/W_2$ in Tables\,\ref{met_table_JK} and \ref{met_table_HK}) for all the colour bins within the uncertainties (although they are relatively large for the colour bins dominated by the MWNSD, where the number of stars per colour bin is smaller). This means that metal rich stars dominate all the analysed colour bins and that the fraction between the metal poor and rich stars does not significantly vary with extinction and/or distance from Sgr\,A*.

We speculate that the detected metal poor and rich components might have different origins, which is in agreement with the slower rotation profile found for the metal poor stars in comparison to the metal rich ones in the MWNSD \citep{Schultheis:2021wf}. Metal rich stars show a rotation which agrees with the gas motion in the central molecular zone suggesting in-situ formation, whereas metal poor stars might have been accreted from stellar clusters contributing to building the MWNSC and the MWNSD. This is in agreement with previous work proposing a hybrid formation scenario for the MWNSC \citep[e.g.][]{Guillard:2016aa,van-Donkelaar:2023aa} combining in-situ star formation \citep[e.g.][]{Milosavljevic:2004bh,McLaughlin:2006fk,Bekki:2007aa} and accretion of star clusters \citep[e.g.][]{tremaine75,Capuzzo-Dolcetta:1993aa,Antonini:2012aa,Antonini:2013ys,Perets:2014aa,Gnedin:2014fk,Arca-Sedda:2015aa,Tsatsi:2017aa,Arca-Sedda:2020ts}. In the framework of this model, we could interpret the metal poor stars in the MWNSC and MWNSD as stars accreted from clusters within a radius of 1.5\,kpc at $z>4$ \citep{van-Donkelaar:2023aa}.

If a significant part of the MWNSC mass originated from accretion of stellar clusters, we would expect that some of these clusters were destroyed before getting to the centre, whereas others could have lost a significant part of their mass in the MWNSD, potentially originating the detected smooth transition between both Galactic centre components. Alternatively, recent simulations have shown that in the case of a galaxy merger, part of the stellar population of the MWNSD might come from the NSCs of the progenitor galaxies and thus part of the MWNSD would share a common origin with the MWNSC \citep{Mastrobuono-Battisti:2023aa}, which can also contribute to the detected smooth transition between the MWNSD and the MWNSC.

The current paradigm for the formation of NSDs in barred spiral galaxies assumes that gas from the galactic disc is funnelled towards the innermost regions of the galaxy by a bar, growing NSDs from inside out and generating a gradient in the properties of their stellar population \citep[e.g.][]{Shlosman:2001aa,Sormani:2015aa,Seo:2019aa,Bittner:2020aa,de-Sa-Freitas:2023aa}. This is compatible with the detected kinematic and metallicity gradients and also with previous results on the MWNSD stellar population which show that this mechanism is also at play in the Milky Way \citep[e.g.][]{Nogueras-Lara:2023aa}.


\subsection{Metallicity gradients in other clusters}

Metallicity gradients resembling the one that we detect in the MWNSC have also been found in other NSCs \citep[e.g. M54, NGC 5102, or NGC 5206, see ][]{Siegel:2007aa,Hannah:2021aa}. They were explained as the combination of a young (several Gyr old) and metal rich stellar population concentrated towards the central regions of the clusters, and an older and metal poorer stellar population which extends to larger radii. Although a direct comparison with the detected gradient in the MWNSC is hard, this mechanism can also contribute to the detected metallicity gradient, given the presence of a young stellar population (several Myr old) in the central region of the MWNSC \citep{Pfuhl:2011uq,Nishiyama:2016zr,Schodel:2020aa}. This scenario is in agreement with previous simulations showing that even evolved NSCs with in situ star formation can contain stellar populations segregated by age \citep{Aharon:2015kx}. Particularly, \citet{Mastrobuono-Battisti:2019aa} found that MWNSC stellar populations born at different epochs present different morphologies and kinematics, being the oldest stars (presumably metal poor) more relaxed and extended \citep{Mastrobuono-Battisti:2019aa}.


\subsection{Different components of the same structure}

Our results open the possibility that the MWNSC and the MWNSD could be part of the same structure with the MWNSD being an extended wing of the MWNSC. In this scenario, previous work showing different stellar properties between these components such as different kinematics and metallicity \citep[e.g.][]{Schultheis:2021wf,Nogueras-Lara:2022tp}, would correspond to partial results that only refer to average quantities obtained in regions dominated by either the MWNSC or the MWNSD, without considering the transition between them. In this way, the detected metallicity gradient also agrees with the results obtained by \citet{Feldmeier-Krause:2022vm} for two regions located at $\sim20$\,pc to the Galactic east and west of the MWNSC (corresponding to the transition region between the MWNSD and the MWNSC), where the metallicity was found to be lower than that of the MWNSC.

Regarding stellar populations, the bulk of stars in both Galactic centre components was found to be old \citep[$\gtrsim 8$\,Gyr, e.g.][]{Nogueras-Lara:2019ad,Schodel:2020aa,Nogueras-Lara:2021wm}, which is compatible with a potential common origin. Subsequent different star formation histories do not pose any problem for this scenario because different star forming processes can be operating on different scales and distance from the Galactic centre. In this way, the MWNSD might have grown from the MWNSC and then, the evolution of the Galactic bar and its efficiency at transporting gas towards the Galactic centre \citep[e.g.][]{Bittner:2020aa,Nogueras-Lara:2023aa}, might have produced different stellar populations in the MWNSD and the MWNSC. Moreover, 
in-situ star formation and infall of stellar cluster in the MWNSC might explain the different evolution of their star formation histories \citep[e.g.][]{Pfuhl:2011uq,Feldmeier-Krause:2020uv}, after a plausible common formation.

The possibility that the MWNSC and the MWNSD are part of the same structure might also explain the detection of very large NSCs \citep[such as the one in FCC\,47, with an effective radius of $\sim70$\,pc,][]{Fahrion:2019aa}, where maybe the two components are present and indistinguishable. Although NSCs are well studied structures \citep[e.g.][]{Neumayer:2020aa}, this is not the case for NSDs which have only been analysed for several tens of galaxies \citep[e.g.][]{Pizzella:2002aa,Ledo:2010aa,de-Lorenzo-Caceres:2019aa,Gadotti:2020aa}. In this way, to shed light on the relation between NSCs and NSDs in general, it is imperative to conduct a statistical study of the coexistence of these two structures in external galaxies as well as a characterisation of their stellar properties.

To further investigate this question, it is necessary to analyse a larger field of view and also obtain more accurate proper motions and line-of-sight velocities to enable an orbit analysis that will allow us to better characterise the transition between the MWNSD and the MWNSC. Moreover, deeper data along the line of sight will  also allow us to improve the study of the stellar population from the backside of the MWNSC and the MWNSD.

\section{Conclusions}

We studied the stellar population along the line of sight towards the MWNSC in a field of $\sim2.8'\times4.9'$ centred on the MWNSC,  to better characterise the relation between the MWNSD and the MWNSC. We used a photometric survey covering the region \citep{Nogueras-Lara:2018aa,Nogueras-Lara:2019aa}, a proper motion catalogue \citep{Shahzamanian:2021wu}, and an independent spectroscopic survey of $\sim1000$ stars with photometric counterparts in the analysed region, and measured $[M/H]$ and line-of-sight velocities \citep{Feldmeier-Krause:2017kq,Feldmeier-Krause:2020uv}.

We applied colour-cuts to independently analyse stars with different extinctions, and distinguish between stars belonging to the MWNSD and the MWNSC \citep{Nogueras-Lara:2021wm,Nogueras-Lara:2022tp}. We detected a gradient in the proper motion component parallel to the Galactic plane which indicates that there is a correlation between colour (i.e. extinction) and distance from the Galactic centre. We found a smooth kinematic transition between regions dominated by the MWNSD and the MWNSC. We confirmed the correlation between extinction and distance to the Galactic centre by analysing the line-of-sight velocity dispersion of spectroscopically characterised stars, and found that the velocity dispersion is largest for stars with $J-K_s\sim7$\,mag and $H-K_s\sim2.5$\,mag, supporting that they are on average closest to Sgr\,A* \citep[e.g.][]{Trippe:2008it,Schodel:2009zr}.

We studied the metallicity distribution of the stars in the field for the previously defined colour cuts. We concluded that there is a bimodal metallicity distribution for all the colour bins, in agreement with previous results on the bulk population of the MWNSD and the MWNSC \citep{Nogueras-Lara:2022tp,Schultheis:2021wf}. We found that the ratio between the metal poor and rich stars is approximately constant within the uncertainties for all the colour bins, indicating that the relative fraction between metal poor and rich stars is not affected by extinction and/or distance from Sgr\,A*. This suggests that the metallicity bi-modality found in previous work when analysing the bulk population of the MWNSC and the MWNSD \citep{Nogueras-Lara:2022tp,Schultheis:2021wf} is a real feature of the MWNSD and the MWNSC. We  speculate that metal rich stars might have formed in-situ from gas transported by the Galactic bar \citep{Schultheis:2021wf,Bittner:2020aa}, whereas metal poor stars might have been accreted from stellar clusters contributing to building the MWNSD and the MWNSC \citep{Antonini:2012aa,Tsatsi:2017aa,van-Donkelaar:2023aa}.

We also analysed the variation of metallicity with extinction for the MWNSD and MWNSC metal poor and rich stars  and detected a metallicity gradient for both stellar populations towards the innermost regions of the MWNSC. This metallicity gradient agrees with the results obtained for two regions $\sim20$\,pc to the Galactic east and west of the MWNSC, where the metallicity was found to decrease in comparison with the values found for the MWNSC  \citep{Feldmeier-Krause:2022vm}. Moreover, similar metallicity gradients have been found in the central regions of nearby external barred spiral galaxies \citep{Gadotti:2020aa,Bittner:2020aa,de-Sa-Freitas:2023aa}, pointing towards an inside-out growth of these nuclear stellar structures and indicating that they are connecting to the galactic bar.

The detected kinematic and metallicity gradients cannot be explained by the mixture of stars from the MWNSD and the MWNSC, and thus an intrinsic variation of their properties seems to be the most plausible explanation. To further constrain our results a full kinematic characterisation of the stars in the transition region between the MWNSD and the MWNSC is necessary to assign membership probabilities, and determine the origin of stars with different metallicities. Our results suggest a smooth transition between the MWNSD and the MWNSC and open the possibility that they might be part of the same structure with the MWNSD being the grown edge of the MWNSC.

     
\begin{acknowledgements}
  
We thank the anonymous referee for helpful comments and suggestions that improved this manuscript. This work is based on observations made with ESO Telescopes at the La Silla Paranal Observatory under program IDs 60.A-9450(A), 091.B-0418, 093.B-0368, and 195.B-0283. FN-L gratefully acknowledges the sponsorship provided by the European Southern Observatory through a research fellowship. RS acknowledges financial support from the State Agency for Research of the Spanish MCIU through the “Center of Excellence Severo Ochoa” award for the Instituto de Astrofísica de Andalucía (SEV-2017-0709) and from grant EUR2022-134031 funded by MCIN/AEI/
10.13039/501100011033 and  by the European Union NextGenerationEU/PRTR. AdLC acknowledges support from the Agencia Estatal de Investigación del Ministerio de Ciencia e Innovación (MCIN/AEI) under grant CoBEARD and the European Regional Development Fund (ERDF) with reference PID2021-128131NB-I00. AdLC also acknowledges financial support from MCIN through the Spanish State Research Agency, under the Severo Ochoa Centres of Excellence Programme 2020-2023 (CEX2019-000920-S), as well as from grant AYA2016-77237-C3-1-P from the Spanish Ministry of Economy and Competitiveness (MINECO).

\end{acknowledgements}

\bibliography{../../BibGC.bib}

\begin{thebibliography}{87}
\expandafter\ifx\csname natexlab\endcsname\relax\def\natexlab#1{#1}\fi

\bibitem[{{Aharon} \& {Perets}(2015)}]{Aharon:2015kx}
{Aharon}, D. \& {Perets}, H.~B. 2015, \apj, 799, 185

\bibitem[{Akaike(1974)}]{Akaike:1974aa}
Akaike, H. 1974, Automatic Control, IEEE Transactions on, 19, 716

\bibitem[{{Antonini}(2013)}]{Antonini:2013ys}
{Antonini}, F. 2013, \apj, 763, 62

\bibitem[{{Antonini} {et~al.}(2012){Antonini}, {Capuzzo-Dolcetta},
  {Mastrobuono-Battisti}, \& {Merritt}}]{Antonini:2012aa}
{Antonini}, F., {Capuzzo-Dolcetta}, R., {Mastrobuono-Battisti}, A., \&
  {Merritt}, D. 2012, \apj, 750, 111

\bibitem[{{Arca-Sedda} {et~al.}(2015){Arca-Sedda}, {Capuzzo-Dolcetta},
  {Antonini}, \& {Seth}}]{Arca-Sedda:2015aa}
{Arca-Sedda}, M., {Capuzzo-Dolcetta}, R., {Antonini}, F., \& {Seth}, A. 2015,
  \apj, 806, 220

\bibitem[{{Arca Sedda} {et~al.}(2020){Arca Sedda}, {Gualandris}, {Do},
  {Feldmeier-Krause}, {Neumayer}, \& {Erkal}}]{Arca-Sedda:2020ts}
{Arca Sedda}, M., {Gualandris}, A., {Do}, T., {et~al.} 2020, \apjl, 901, L29

\bibitem[{{Bekki}(2007)}]{Bekki:2007aa}
{Bekki}, K. 2007, \pasa, 24, 77

\bibitem[{{Bittner} {et~al.}(2020){Bittner}, {S{\'a}nchez-Bl{\'a}zquez},
  {Gadotti}, {Neumann}, {Fragkoudi}, {Coelho}, {de Lorenzo-C{\'a}ceres},
  {Falc{\'o}n-Barroso}, {Kim}, {Leaman}, {Mart{\'\i}n-Navarro},
  {M{\'e}ndez-Abreu}, {P{\'e}rez}, {Querejeta}, {Seidel}, \& {van de
  Ven}}]{Bittner:2020aa}
{Bittner}, A., {S{\'a}nchez-Bl{\'a}zquez}, P., {Gadotti}, D.~A., {et~al.} 2020,
  \aap, 643, A65

\bibitem[{{Capuzzo-Dolcetta}(1993)}]{Capuzzo-Dolcetta:1993aa}
{Capuzzo-Dolcetta}, R. 1993, \apj, 415, 616

\bibitem[{{Chatzopoulos} {et~al.}(2015{\natexlab{a}}){Chatzopoulos}, {Fritz},
  {Gerhard}, {Gillessen}, {Wegg}, {Genzel}, \& {Pfuhl}}]{Chatzopoulos:2015yu}
{Chatzopoulos}, S., {Fritz}, T.~K., {Gerhard}, O., {et~al.} 2015{\natexlab{a}},
  \mnras, 447, 948

\bibitem[{{Chatzopoulos} {et~al.}(2015{\natexlab{b}}){Chatzopoulos}, {Gerhard},
  {Fritz}, {Wegg}, {Gillessen}, {Pfuhl}, \& {Eisenhauer}}]{Chatzopoulos:2015uq}
{Chatzopoulos}, S., {Gerhard}, O., {Fritz}, T.~K., {et~al.} 2015{\natexlab{b}},
  \mnras, 453, 939

\bibitem[{{Chen} {et~al.}(2023){Chen}, {Do}, {Ghez}, {Hosek},
  {Feldmeier-Krause}, {Chu}, {Bentley}, {Lu}, \& {Morris}}]{Chen:2023aa}
{Chen}, Z., {Do}, T., {Ghez}, A.~M., {et~al.} 2023, \apj, 944, 79

\bibitem[{{de Lorenzo-C{\'a}ceres} {et~al.}(2019){de Lorenzo-C{\'a}ceres},
  {S{\'a}nchez-Bl{\'a}zquez}, {M{\'e}ndez-Abreu}, {Gadotti},
  {Falc{\'o}n-Barroso}, {Mart{\'\i}nez-Valpuesta}, {Coelho}, {Fragkoudi},
  {Husemann}, {Leaman}, {P{\'e}rez}, {Querejeta}, {Seidel}, \& {van de
  Ven}}]{de-Lorenzo-Caceres:2019aa}
{de Lorenzo-C{\'a}ceres}, A., {S{\'a}nchez-Bl{\'a}zquez}, P.,
  {M{\'e}ndez-Abreu}, J., {et~al.} 2019, \mnras, 484, 5296

\bibitem[{{de S{\'a}-Freitas} {et~al.}(2023){de S{\'a}-Freitas}, {Fragkoudi},
  {Gadotti}, {Falc{\'o}n-Barroso}, {Bittner}, {S{\'a}nchez-Bl{\'a}zquez}, {van
  de Ven}, {Bieri}, {Coccato}, {Coelho}, {Fahrion}, {Gon{\c{c}}alves}, {Kim},
  {de Lorenzo-C{\'a}ceres}, {Martig}, {Mart{\'\i}n-Navarro}, {Mendez-Abreu},
  {Neumann}, \& {Querejeta}}]{de-Sa-Freitas:2023aa}
{de S{\'a}-Freitas}, C., {Fragkoudi}, F., {Gadotti}, D.~A., {et~al.} 2023,
  \aap, 671, A8

\bibitem[{{Do} {et~al.}(2020){Do}, {David Martinez}, {Kerzendorf},
  {Feldmeier-Krause}, {Arca Sedda}, {Neumayer}, \& {Gualandris}}]{Do:2020wr}
{Do}, T., {David Martinez}, G., {Kerzendorf}, W., {et~al.} 2020, \apjl, 901,
  L28

\bibitem[{{Do} {et~al.}(2019){Do}, {Hees}, {Ghez}, {Martinez}, {Chu}, {Jia},
  {Sakai}, {Lu}, {Gautam}, {O{\textquoteright}Neil}, {Becklin}, {Morris},
  {Matthews}, {Nishiyama}, {Campbell}, {Chappell}, {Chen}, {Ciurlo},
  {Dehghanfar}, {Gallego-Cano}, {Kerzendorf}, {Lyke}, {Naoz}, {Saida},
  {Sch{\"o}del}, {Takahashi}, {Takamori}, {Witzel}, \&
  {Wizinowich}}]{Do:2019aa}
{Do}, T., {Hees}, A., {Ghez}, A., {et~al.} 2019, Science, 365, 664

\bibitem[{{Dong} {et~al.}(2011){Dong}, {Wang}, {Cotera}, {Stolovy}, {Morris},
  {Mauerhan}, {Mills}, {Schneider}, {Calzetti}, \& {Lang}}]{Dong:2011ff}
{Dong}, H., {Wang}, Q.~D., {Cotera}, A., {et~al.} 2011, \mnras, 417, 114

\bibitem[{{Fahrion} {et~al.}(2019){Fahrion}, {Lyubenova}, {van de Ven},
  {Leaman}, {Hilker}, {Mart{\'\i}n-Navarro}, {Zhu}, {Alfaro-Cuello}, {Coccato},
  {Corsini}, {Falc{\'o}n-Barroso}, {Iodice}, {McDermid}, {Sarzi}, \& {de
  Zeeuw}}]{Fahrion:2019aa}
{Fahrion}, K., {Lyubenova}, M., {van de Ven}, G., {et~al.} 2019, \aap, 628, A92

\bibitem[{{Feldmeier} {et~al.}(2014){Feldmeier}, {Neumayer}, {Seth},
  {Sch{\"o}del}, {L{\"u}tzgendorf}, {de Zeeuw}, {Kissler-Patig}, {Nishiyama},
  \& {Walcher}}]{Feldmeier:2014kx}
{Feldmeier}, A., {Neumayer}, N., {Seth}, A., {et~al.} 2014, \aap, 570, A2

\bibitem[{{Feldmeier-Krause}(2022)}]{Feldmeier-Krause:2022vm}
{Feldmeier-Krause}, A. 2022, \mnras, 513, 5920

\bibitem[{{Feldmeier-Krause} {et~al.}(2020){Feldmeier-Krause}, {Kerzendorf},
  {Do}, {Nogueras-Lara}, {Neumayer}, {Walcher}, {Seth}, {Sch{\"o}del}, {de
  Zeeuw}, {Hilker}, {L{\"u}tzgendorf}, {Kuntschner}, \&
  {Kissler-Patig}}]{Feldmeier-Krause:2020uv}
{Feldmeier-Krause}, A., {Kerzendorf}, W., {Do}, T., {et~al.} 2020, \mnras, 494,
  396

\bibitem[{{Feldmeier-Krause} {et~al.}(2017{\natexlab{a}}){Feldmeier-Krause},
  {Kerzendorf}, {Neumayer}, {Sch{\"o}del}, {Nogueras-Lara}, {Do}, {de Zeeuw},
  \& {Kuntschner}}]{Feldmeier-Krause:2017kq}
{Feldmeier-Krause}, A., {Kerzendorf}, W., {Neumayer}, N., {et~al.}
  2017{\natexlab{a}}, \mnras, 464, 194

\bibitem[{{Feldmeier-Krause} {et~al.}(2017{\natexlab{b}}){Feldmeier-Krause},
  {Zhu}, {Neumayer}, {van de Ven}, {de Zeeuw}, \&
  {Sch{\"o}del}}]{Feldmeier-Krause:2017tk}
{Feldmeier-Krause}, A., {Zhu}, L., {Neumayer}, N., {et~al.} 2017{\natexlab{b}},
  \mnras, 466, 4040

\bibitem[{{Fritz} {et~al.}(2016){Fritz}, {Chatzopoulos}, {Gerhard},
  {Gillessen}, {Genzel}, {Pfuhl}, {Tacchella}, {Eisenhauer}, \&
  {Ott}}]{Fritz:2016aa}
{Fritz}, T.~K., {Chatzopoulos}, S., {Gerhard}, O., {et~al.} 2016, \apj, 821, 44

\bibitem[{{Fritz} {et~al.}(2011){Fritz}, {Gillessen}, {Dodds-Eden}, {Lutz},
  {Genzel}, {Raab}, {Ott}, {Pfuhl}, {Eisenhauer}, \&
  {Yusef-Zadeh}}]{Fritz:2011fk}
{Fritz}, T.~K., {Gillessen}, S., {Dodds-Eden}, K., {et~al.} 2011, \apj, 737, 73

\bibitem[{{Fritz} {et~al.}(2021){Fritz}, {Patrick}, {Feldmeier-Krause},
  {Sch{\"o}del}, {Schultheis}, {Gerhard}, {Nandakumar}, {Neumayer},
  {Nogueras-Lara}, \& {Prieto}}]{Fritz:2020aa}
{Fritz}, T.~K., {Patrick}, L.~R., {Feldmeier-Krause}, A., {et~al.} 2021, \aap,
  649, A83

\bibitem[{{Gadotti} {et~al.}(2020){Gadotti}, {Bittner}, {Falc{\'o}n-Barroso},
  {M{\'e}ndez-Abreu}, {Kim}, {Fragkoudi}, {de Lorenzo-C{\'a}ceres}, {Leaman},
  {Neumann}, {Querejeta}, {S{\'a}nchez-Bl{\'a}zquez}, {Martig},
  {Mart{\'\i}n-Navarro}, {P{\'e}rez}, {Seidel}, \& {van de
  Ven}}]{Gadotti:2020aa}
{Gadotti}, D.~A., {Bittner}, A., {Falc{\'o}n-Barroso}, J., {et~al.} 2020, \aap,
  643, A14

\bibitem[{{Gallego-Cano} {et~al.}(2020){Gallego-Cano}, {Sch{\"o}del},
  {Nogueras-Lara}, {Dong}, {Shahzamanian}, {Fritz}, {Gallego-Calvente}, \&
  {Neumayer}}]{gallego-cano2019}
{Gallego-Cano}, E., {Sch{\"o}del}, R., {Nogueras-Lara}, F., {et~al.} 2020,
  \aap, 634, A71

\bibitem[{{Girardi}(2016)}]{Girardi:2016fk}
{Girardi}, L. 2016, \araa, 54, 95

\bibitem[{{Gnedin} {et~al.}(2014){Gnedin}, {Ostriker}, \&
  {Tremaine}}]{Gnedin:2014fk}
{Gnedin}, O.~Y., {Ostriker}, J.~P., \& {Tremaine}, S. 2014, \apj, 785, 71

\bibitem[{{Gonzalez} {et~al.}(2012){Gonzalez}, {Rejkuba}, {Zoccali}, {Valenti},
  {Minniti}, {Schultheis}, {Tobar}, \& {Chen}}]{Gonzalez:2012aa}
{Gonzalez}, O.~A., {Rejkuba}, M., {Zoccali}, M., {et~al.} 2012, \aap, 543, A13

\bibitem[{{Gravity Collaboration} {et~al.}(2018){Gravity Collaboration},
  {Abuter}, {Amorim}, {Anugu}, {Baub{\"o}ck}, {Benisty}, {Berger}, {Blind},
  {Bonnet}, {Brandner}, {Buron}, {Collin}, {Chapron}, {Cl{\'e}net}, {Coud{\'e}
  Du Foresto}, {de Zeeuw}, {Deen}, {Delplancke-Str{\"o}bele}, {Dembet},
  {Dexter}, {Duvert}, {Eckart}, {Eisenhauer}, {Finger}, {F{\"o}rster
  Schreiber}, {F{\'e}dou}, {Garcia}, {Garcia Lopez}, {Gao}, {Gendron},
  {Genzel}, {Gillessen}, {Gordo}, {Habibi}, {Haubois}, {Haug}, {Hau{\ss}mann},
  {Henning}, {Hippler}, {Horrobin}, {Hubert}, {Hubin}, {Jimenez Rosales},
  {Jochum}, {Jocou}, {Kaufer}, {Kellner}, {Kendrew}, {Kervella}, {Kok},
  {Kulas}, {Lacour}, {Lapeyr{\`e}re}, {Lazareff}, {Le Bouquin}, {L{\'e}na},
  {Lippa}, {Lenzen}, {M{\'e}rand}, {M{\"u}ler}, {Neumann}, {Ott}, {Palanca},
  {Paumard}, {Pasquini}, {Perraut}, {Perrin}, {Pfuhl}, {Plewa}, {Rabien},
  {Ram{\'{\i}}rez}, {Ramos}, {Rau}, {Rodr{\'{\i}}guez-Coira}, {Rohloff},
  {Rousset}, {Sanchez-Bermudez}, {Scheithauer}, {Sch{\"o}ller}, {Schuler},
  {Spyromilio}, {Straub}, {Straubmeier}, {Sturm}, {Tacconi}, {Tristram},
  {Vincent}, {von Fellenberg}, {Wank}, {Waisberg}, {Widmann}, {Wieprecht},
  {Wiest}, {Wiezorrek}, {Woillez}, {Yazici}, {Ziegler}, \&
  {Zins}}]{Gravity-Collaboration:2018aa}
{Gravity Collaboration}, {Abuter}, R., {Amorim}, A., {et~al.} 2018, \aap, 615,
  L15

\bibitem[{{Guillard} {et~al.}(2016){Guillard}, {Emsellem}, \&
  {Renaud}}]{Guillard:2016aa}
{Guillard}, N., {Emsellem}, E., \& {Renaud}, F. 2016, \mnras, 461, 3620

\bibitem[{{Hannah} {et~al.}(2021){Hannah}, {Seth}, {Nguyen}, {Dumont},
  {Kacharov}, {Neumayer}, \& {den Brok}}]{Hannah:2021aa}
{Hannah}, C.~H., {Seth}, A.~C., {Nguyen}, D.~D., {et~al.} 2021, \aj, 162, 281

\bibitem[{{Husser} {et~al.}(2013){Husser}, {Wende-von Berg}, {Dreizler},
  {Homeier}, {Reiners}, {Barman}, \& {Hauschildt}}]{Husser:2013uu}
{Husser}, T.~O., {Wende-von Berg}, S., {Dreizler}, S., {et~al.} 2013, \aap,
  553, A6

\bibitem[{{Kerzendorf} \& {Do}(2015)}]{Kerzendorf:2015aa}
{Kerzendorf}, W. \& {Do}, T. 2015

\bibitem[{{Kissler-Patig} {et~al.}(2008){Kissler-Patig}, {Pirard}, {Casali},
  {Moorwood}, {Ageorges}, {Alves de Oliveira}, {Baksai}, {Bedin}, {Bendek},
  {Biereichel}, {Delabre}, {Dorn}, {Esteves}, {Finger}, {Gojak}, {Huster},
  {Jung}, {Kiekebush}, {Klein}, {Koch}, {Lizon}, {Mehrgan}, {Petr-Gotzens},
  {Pritchard}, {Selman}, \& {Stegmeier}}]{Kissler-Patig:2008uq}
{Kissler-Patig}, M., {Pirard}, J.-F., {Casali}, M., {et~al.} 2008, \aap, 491,
  941

\bibitem[{{Launhardt} {et~al.}(2002){Launhardt}, {Zylka}, \&
  {Mezger}}]{Launhardt:2002nx}
{Launhardt}, R., {Zylka}, R., \& {Mezger}, P.~G. 2002, \aap, 384, 112

\bibitem[{Ledo {et~al.}(2010)Ledo, Sarzi, Dotti, Khochfar, \&
  Morelli}]{Ledo:2010aa}
Ledo, H.~R., Sarzi, M., Dotti, M., Khochfar, S., \& Morelli, L. 2010, Monthly
  Notices of the Royal Astronomical Society, 407, 969

\bibitem[{{Lyubenova} {et~al.}(2013){Lyubenova}, {van den Bosch},
  {C{\^o}t{\'e}}, {Kuntschner}, {van de Ven}, {Ferrarese}, {Jord{\'a}n},
  {Infante}, \& {Peng}}]{Lyubenova:2013aa}
{Lyubenova}, M., {van den Bosch}, R. C.~E., {C{\^o}t{\'e}}, P., {et~al.} 2013,
  \mnras, 431, 3364

\bibitem[{{Mart{\'\i}nez-Arranz} {et~al.}(2022){Mart{\'\i}nez-Arranz},
  {Sch{\"o}del}, {Nogueras-Lara}, \& {Shahzamanian}}]{Martinez-Arranz:2022uf}
{Mart{\'\i}nez-Arranz}, {\'A}., {Sch{\"o}del}, R., {Nogueras-Lara}, F., \&
  {Shahzamanian}, B. 2022, \aap, 660, L3

\bibitem[{{Mastrobuono-Battisti} {et~al.}(2023){Mastrobuono-Battisti}, {Ogiya},
  {Hahn}, \& {Schultheis}}]{Mastrobuono-Battisti:2023aa}
{Mastrobuono-Battisti}, A., {Ogiya}, G., {Hahn}, O., \& {Schultheis}, M. 2023,
  \mnras, 521, 6089

\bibitem[{{Mastrobuono-Battisti} {et~al.}(2019){Mastrobuono-Battisti},
  {Perets}, {Gualandris}, {Neumayer}, \&
  {Sippel}}]{Mastrobuono-Battisti:2019aa}
{Mastrobuono-Battisti}, A., {Perets}, H.~B., {Gualandris}, A., {Neumayer}, N.,
  \& {Sippel}, A.~C. 2019, \mnras, 490, 5820

\bibitem[{{McLaughlin} {et~al.}(2006){McLaughlin}, {Anderson}, {Meylan},
  {Gebhardt}, {Pryor}, {Minniti}, \& {Phinney}}]{McLaughlin:2006fk}
{McLaughlin}, D.~E., {Anderson}, J., {Meylan}, G., {et~al.} 2006, \apjs, 166,
  249

\bibitem[{{Milosavljevi{\'c}} \& {Loeb}(2004)}]{Milosavljevic:2004bh}
{Milosavljevi{\'c}}, M. \& {Loeb}, A. 2004, \apjl, 604, L45

\bibitem[{{Nagayama} {et~al.}(2003){Nagayama}, {Nagashima}, {Nakajima},
  {Nagata}, {Sato}, {Nakaya}, {Yamamuro}, {Sugitani}, \&
  {Tamura}}]{Nagayama:2003fk}
{Nagayama}, T., {Nagashima}, C., {Nakajima}, Y., {et~al.} 2003, in \procspie,
  Vol. 4841, Instrument Design and Performance for Optical/Infrared
  Ground-based Telescopes, ed. M.~{Iye} \& A.~F.~M. {Moorwood}, 459--464

\bibitem[{{Neumayer} {et~al.}(2020){Neumayer}, {Seth}, \&
  {B{\"o}ker}}]{Neumayer:2020aa}
{Neumayer}, N., {Seth}, A., \& {B{\"o}ker}, T. 2020, \aapr, 28, 4

\bibitem[{{Nishiyama} {et~al.}(2006){Nishiyama}, {Nagata}, {Kusakabe},
  {Matsunaga}, {Naoi}, {Kato}, {Nagashima}, {Sugitani}, {Tamura}, {Tanab{\'e}},
  \& {Sato}}]{Nishiyama:2006tx}
{Nishiyama}, S., {Nagata}, T., {Kusakabe}, N., {et~al.} 2006, \apj, 638, 839

\bibitem[{{Nishiyama} {et~al.}(2008){Nishiyama}, {Nagata}, {Tamura}, {Kandori},
  {Hatano}, {Sato}, \& {Sugitani}}]{Nishiyama:2008qa}
{Nishiyama}, S., {Nagata}, T., {Tamura}, M., {et~al.} 2008, \apj, 680, 1174

\bibitem[{{Nishiyama} {et~al.}(2016){Nishiyama}, {Sch{\"o}del}, {Yoshikawa},
  {Nagata}, {Minowa}, \& {Tamura}}]{Nishiyama:2016zr}
{Nishiyama}, S., {Sch{\"o}del}, R., {Yoshikawa}, T., {et~al.} 2016, \aap, 588,
  A49

\bibitem[{{Nogueras-Lara}(2022{\natexlab{a}})}]{Nogueras-Lara:2022aa}
{Nogueras-Lara}, F. 2022{\natexlab{a}}, \aap, 668, L8

\bibitem[{{Nogueras-Lara}(2022{\natexlab{b}})}]{Nogueras-Lara:2022tp}
{Nogueras-Lara}, F. 2022{\natexlab{b}}, \aap, 666, A72

\bibitem[{{Nogueras-Lara} {et~al.}(2018{\natexlab{a}}){Nogueras-Lara},
  {Gallego-Calvente}, {Dong}, {Gallego-Cano}, {Girard}, {Hilker}, {de Zeeuw},
  {Feldmeier-Krause}, {Nishiyama}, {Najarro}, {Neumayer}, \&
  {Sch{\"o}del}}]{Nogueras-Lara:2018aa}
{Nogueras-Lara}, F., {Gallego-Calvente}, A.~T., {Dong}, H., {et~al.}
  2018{\natexlab{a}}, \aap, 610, A83

\bibitem[{{Nogueras-Lara} {et~al.}(2018{\natexlab{b}}){Nogueras-Lara},
  {Sch{\"o}del}, {Dong}, {Najarro}, {Gallego-Calvente}, {Hilker},
  {Gallego-Cano}, {Nishiyama}, {Neumayer}, {Feldmeier-Krause}, {Girard},
  {Cassisi}, \& {Pietrinferni}}]{Nogueras-Lara:2018ab}
{Nogueras-Lara}, F., {Sch{\"o}del}, R., {Dong}, H., {et~al.}
  2018{\natexlab{b}}, \aap, 620, A83

\bibitem[{{Nogueras-Lara} {et~al.}(2019){Nogueras-Lara}, {Sch{\"o}del},
  {Gallego-Calvente}, {Dong}, {Gallego-Cano}, {Shahzamanian}, {Girard},
  {Nishiyama}, {Najarro}, \& {Neumayer}}]{Nogueras-Lara:2019aa}
{Nogueras-Lara}, F., {Sch{\"o}del}, R., {Gallego-Calvente}, A.~T., {et~al.}
  2019, \aap, 631, A20

\bibitem[{{Nogueras-Lara} {et~al.}(2020{\natexlab{a}}){Nogueras-Lara},
  {Sch{\"o}del}, {Gallego-Calvente}, {Gallego-Cano}, {Shahzamanian}, {Dong},
  {Neumayer}, {Hilker}, {Najarro}, {Nishiyama}, {Feldmeier-Krause}, {Girard},
  \& {Cassisi}}]{Nogueras-Lara:2019ad}
{Nogueras-Lara}, F., {Sch{\"o}del}, R., {Gallego-Calvente}, A.~T., {et~al.}
  2020{\natexlab{a}}, Nature Astronomy, 4, 377

\bibitem[{{Nogueras-Lara} {et~al.}(2021{\natexlab{a}}){Nogueras-Lara},
  {Sch{\"o}del}, \& {Neumayer}}]{Nogueras-Lara:2021uz}
{Nogueras-Lara}, F., {Sch{\"o}del}, R., \& {Neumayer}, N. 2021{\natexlab{a}},
  \aap, 653, A33

\bibitem[{{Nogueras-Lara} {et~al.}(2021{\natexlab{b}}){Nogueras-Lara},
  {Sch{\"o}del}, \& {Neumayer}}]{Nogueras-Lara:2021wj}
{Nogueras-Lara}, F., {Sch{\"o}del}, R., \& {Neumayer}, N. 2021{\natexlab{b}},
  \aap, 653, A133

\bibitem[{{Nogueras-Lara} {et~al.}(2021{\natexlab{c}}){Nogueras-Lara},
  {Sch{\"o}del}, \& {Neumayer}}]{Nogueras-Lara:2021wm}
{Nogueras-Lara}, F., {Sch{\"o}del}, R., \& {Neumayer}, N. 2021{\natexlab{c}},
  \apj, 920, 97

\bibitem[{{Nogueras-Lara} {et~al.}(2020{\natexlab{b}}){Nogueras-Lara},
  {Sch{\"o}del}, {Neumayer}, {Gallego-Cano}, {Shahzamanian},
  {Gallego-Calvente}, \& {Najarro}}]{Nogueras-Lara:2020aa}
{Nogueras-Lara}, F., {Sch{\"o}del}, R., {Neumayer}, N., {et~al.}
  2020{\natexlab{b}}, \aap, 641, A141

\bibitem[{{Nogueras-Lara} {et~al.}(2023){Nogueras-Lara}, {Schultheis},
  {Najarro}, {Sormani}, {Gadotti}, \& {Rich}}]{Nogueras-Lara:2023aa}
{Nogueras-Lara}, F., {Schultheis}, M., {Najarro}, F., {et~al.} 2023, \aap, 671,
  L10

\bibitem[{Pedregosa {et~al.}(2011)Pedregosa, Varoquaux, Gramfort, Michel,
  Thirion, Grisel, Blondel, Prettenhofer, Weiss, Dubourg, Vanderplas, Passos,
  Cournapeau, Brucher, Perrot, \& Duchesnay}]{Pedregosa:2011aa}
Pedregosa, F., Varoquaux, G., Gramfort, A., {et~al.} 2011, Journal of Machine
  Learning Research, 12, 2825

\bibitem[{{Perets} \& {Mastrobuono-Battisti}(2014)}]{Perets:2014aa}
{Perets}, H.~B. \& {Mastrobuono-Battisti}, A. 2014, \apjl, 784, L44

\bibitem[{{Pfuhl} {et~al.}(2011){Pfuhl}, {Fritz}, {Zilka}, {Maness},
  {Eisenhauer}, {Genzel}, {Gillessen}, {Ott}, {Dodds-Eden}, \&
  {Sternberg}}]{Pfuhl:2011uq}
{Pfuhl}, O., {Fritz}, T.~K., {Zilka}, M., {et~al.} 2011, \apj, 741, 108

\bibitem[{{Pizzella} {et~al.}(2002){Pizzella}, {Corsini}, {Morelli}, {Sarzi},
  {Scarlata}, {Stiavelli}, \& {Bertola}}]{Pizzella:2002aa}
{Pizzella}, A., {Corsini}, E.~M., {Morelli}, L., {et~al.} 2002, \apj, 573, 131

\bibitem[{{Sch{\"o}del} {et~al.}(2014{\natexlab{a}}){Sch{\"o}del}, {Feldmeier},
  {Kunneriath}, {Stolovy}, {Neumayer}, {Amaro-Seoane}, \&
  {Nishiyama}}]{Schodel:2014fk}
{Sch{\"o}del}, R., {Feldmeier}, A., {Kunneriath}, D., {et~al.}
  2014{\natexlab{a}}, \aap, 566, A47

\bibitem[{{Sch{\"o}del} {et~al.}(2014{\natexlab{b}}){Sch{\"o}del}, {Feldmeier},
  {Neumayer}, {Meyer}, \& {Yelda}}]{Schodel:2014bn}
{Sch{\"o}del}, R., {Feldmeier}, A., {Neumayer}, N., {Meyer}, L., \& {Yelda}, S.
  2014{\natexlab{b}}, Classical and Quantum Gravity, 31, 244007

\bibitem[{{Sch{\"o}del} {et~al.}(2009){Sch{\"o}del}, {Merritt}, \&
  {Eckart}}]{Schodel:2009zr}
{Sch{\"o}del}, R., {Merritt}, D., \& {Eckart}, A. 2009, \aap, 502, 91

\bibitem[{{Sch{\"o}del} {et~al.}(2010){Sch{\"o}del}, {Najarro}, {Muzic}, \&
  {Eckart}}]{Schodel:2010fk}
{Sch{\"o}del}, R., {Najarro}, F., {Muzic}, K., \& {Eckart}, A. 2010, \aap, 511,
  A18+

\bibitem[{{Sch{\"o}del} {et~al.}(2020){Sch{\"o}del}, {Nogueras-Lara},
  {Gallego-Cano}, {Shahzamanian}, {Gallego-Calvente}, \&
  {Gardini}}]{Schodel:2020aa}
{Sch{\"o}del}, R., {Nogueras-Lara}, F., {Gallego-Cano}, E., {et~al.} 2020,
  \aap, 641, A102

\bibitem[{{Sch{\"o}del} {et~al.}(2013){Sch{\"o}del}, {Yelda}, {Ghez}, {Girard},
  {Labadie}, {Rebolo}, {P{\'e}rez-Garrido}, \& {Morris}}]{Schodel:2013fk}
{Sch{\"o}del}, R., {Yelda}, S., {Ghez}, A., {et~al.} 2013, \mnras, 429, 1367

\bibitem[{{Schultheis} {et~al.}(2021){Schultheis}, {Fritz}, {Nandakumar},
  {Rojas-Arriagada}, {Nogueras-Lara}, {Feldmeier-Krause}, {Gerhard},
  {Neumayer}, {Patrick}, {Prieto}, {Sch{\"o}del}, {Mastrobuono-Battisti}, \&
  {Sormani}}]{Schultheis:2021wf}
{Schultheis}, M., {Fritz}, T.~K., {Nandakumar}, G., {et~al.} 2021, \aap, 650,
  A191

\bibitem[{{Scoville} {et~al.}(2003){Scoville}, {Stolovy}, {Rieke},
  {Christopher}, \& {Yusef-Zadeh}}]{Scoville:2003la}
{Scoville}, N.~Z., {Stolovy}, S.~R., {Rieke}, M., {Christopher}, M., \&
  {Yusef-Zadeh}, F. 2003, \apj, 594, 294

\bibitem[{{Seo} {et~al.}(2019){Seo}, {Kim}, {Kwak}, {Hsieh}, {Han}, \&
  {Hopkins}}]{Seo:2019aa}
{Seo}, W.-Y., {Kim}, W.-T., {Kwak}, S., {et~al.} 2019, \apj, 872, 5

\bibitem[{{Shahzamanian} {et~al.}(2022){Shahzamanian}, {Sch{\"o}del},
  {Nogueras-Lara}, {Mart{\'\i}nez-Arranz}, {Sormani}, {Gallego-Calvente},
  {Gallego-Cano}, \& {Alburai}}]{Shahzamanian:2021wu}
{Shahzamanian}, B., {Sch{\"o}del}, R., {Nogueras-Lara}, F., {et~al.} 2022,
  \aap, 662, A11

\bibitem[{{Shlosman}(2001)}]{Shlosman:2001aa}
{Shlosman}, I. 2001, 249, 55

\bibitem[{{Siegel} {et~al.}(2007){Siegel}, {Dotter}, {Majewski}, {Sarajedini},
  {Chaboyer}, {Nidever}, {Anderson}, {Mar{\'\i}n-Franch}, {Rosenberg}, {Bedin},
  {Aparicio}, {King}, {Piotto}, \& {Reid}}]{Siegel:2007aa}
{Siegel}, M.~H., {Dotter}, A., {Majewski}, S.~R., {et~al.} 2007, \apjl, 667,
  L57

\bibitem[{{Sormani} {et~al.}(2015){Sormani}, {Binney}, \&
  {Magorrian}}]{Sormani:2015aa}
{Sormani}, M.~C., {Binney}, J., \& {Magorrian}, J. 2015, \mnras, 449, 2421

\bibitem[{{Sormani} {et~al.}(2020){Sormani}, {Magorrian}, {Nogueras-Lara},
  {Neumayer}, {Sch{\"o}nrich}, {Klessen}, \&
  {Mastrobuono-Battisti}}]{Sormani:2020aa}
{Sormani}, M.~C., {Magorrian}, J., {Nogueras-Lara}, F., {et~al.} 2020, \mnras,
  499, 7

\bibitem[{{Sormani} {et~al.}(2022){Sormani}, {Sanders}, {Fritz}, {Smith},
  {Gerhard}, {Sch{\"o}del}, {Magorrian}, {Neumayer}, {Nogueras-Lara},
  {Feldmeier-Krause}, {Mastrobuono-Battisti}, {Schultheis}, {Shahzamanian},
  {Vasiliev}, {Klessen}, {Lucas}, \& {Minniti}}]{Sormani:2022wv}
{Sormani}, M.~C., {Sanders}, J.~L., {Fritz}, T.~K., {et~al.} 2022, \mnras, 512,
  1857

\bibitem[{{Stolovy} {et~al.}(2006){Stolovy}, {Ramirez}, {Arendt}, {Cotera},
  {Yusef-Zadeh}, {Law}, {Gezari}, {Sellgren}, {Karr}, {Moseley}, \&
  {Smith}}]{Stolovy:2006fk}
{Stolovy}, S., {Ramirez}, S., {Arendt}, R.~G., {et~al.} 2006, Journal of
  Physics Conference Series, 54, 176

\bibitem[{{Surot} {et~al.}(2020){Surot}, {Valenti}, {Gonzalez}, {Zoccali},
  {S{\"o}kmen}, {Hidalgo}, \& {Minniti}}]{Surot:2020vo}
{Surot}, F., {Valenti}, E., {Gonzalez}, O.~A., {et~al.} 2020, \aap, 644, A140

\bibitem[{{Tremaine} {et~al.}(1975){Tremaine}, {Ostriker}, \&
  {Spitzer}}]{tremaine75}
{Tremaine}, S.~D., {Ostriker}, J.~P., \& {Spitzer}, Jr., L. 1975, \apj, 196,
  407

\bibitem[{{Trippe} {et~al.}(2008){Trippe}, {Gillessen}, {Gerhard}, {Bartko},
  {Fritz}, {Maness}, {Eisenhauer}, {Martins}, {Ott}, {Dodds-Eden}, \&
  {Genzel}}]{Trippe:2008it}
{Trippe}, S., {Gillessen}, S., {Gerhard}, O.~E., {et~al.} 2008, \aap, 492, 419

\bibitem[{{Tsatsi} {et~al.}(2017){Tsatsi}, {Mastrobuono-Battisti}, {van de
  Ven}, {Perets}, {Bianchini}, \& {Neumayer}}]{Tsatsi:2017aa}
{Tsatsi}, A., {Mastrobuono-Battisti}, A., {van de Ven}, G., {et~al.} 2017,
  \mnras, 464, 3720

\bibitem[{{van Donkelaar} {et~al.}(2023){van Donkelaar}, {Mayer}, {Capelo},
  {Tamfal}, {Quinn}, \& {Madau}}]{van-Donkelaar:2023aa}
{van Donkelaar}, F., {Mayer}, L., {Capelo}, P.~R., {et~al.} 2023, \mnras, 522,
  1726

\bibitem[{{Wang} {et~al.}(2010){Wang}, {Dong}, {Cotera}, {Stolovy}, {Morris},
  {Lang}, {Muno}, {Schneider}, \& {Calzetti}}]{Wang:2010fk}
{Wang}, Q.~D., {Dong}, H., {Cotera}, A., {et~al.} 2010, \mnras, 402, 895

\end{thebibliography}

\appendix

  \section{Proper motion analysis}

\begin{table}

\caption{Median proper motions and their uncertainties for the colour bins in the CMD $K_s$ versus $J-K_s$.}
\label{proper_motions_JK} 
\begin{center}
\def\arraystretch{1.3}
\setlength{\tabcolsep}{1.6pt}    
    
\small
\begin{tabular}{cccccccc}
 &  &  &  &  &  &  & \tabularnewline
\hline 
\hline 
$J-K_s$ & \#stars & $\mu_l$ & $d\mu_l$ & $\sigma\mu_l$ & $\mu_b$ & $d\mu_b$ & $\sigma\mu_b$\tabularnewline
(mag) &  & (mas/yr) & (mas/yr) & (mas/yr) & (mas/yr) & (mas/yr) & (mas/yr)\tabularnewline
\hline 
4-4.5 & 117 & 2.22 & 0.33 & 2.82 & -0.21 & 0.24 & 2.10\tabularnewline
4.5-5 & 170 & 1.64 & 0.27 & 2.79 & 0.11 & 0.23 & 2.35\tabularnewline
5-5.5 & 281 & 0.64 & 0.20 & 2.65 & -0.07 & 0.18 & 2.29\tabularnewline
5.5-6 & 342 & 0.06 & 0.19 & 2.80 & 0.06 & 0.18 & 2.64\tabularnewline
6-6.5 & 339 & 0.03 & 0.20 & 2.90 & -0.02 & 0.18 & 2.65\tabularnewline
6.5-7 & 254 & -0.05 & 0.20 & 2.51 & -0.07 & 0.20 & 2.50\tabularnewline
7-7.5 & 157 & -0.78 & 0.27 & 2.73 & -0.31 & 0.28 & 2.84\tabularnewline
\hline 

\end{tabular}    
    
\end{center}
\footnotesize
\textbf{Notes.} $J-K_s$, $\mu_i$, $d\mu_i$, and $\sigma\mu_i$ indicate the approximate colour range (see Fig.\,\ref{CMD_pm}) and the median value of the proper motion distribution with its corresponding standard error and the standard deviation, respectively. The subindex $i$ indicates Galactic longitude ($i=l$) or latitude ($i=b$).

 \end{table}

\begin{table}

\caption{Median proper motions and their uncertainties for the colour bins in the CMD $K_s$ versus $H-K_s$.}
\label{proper_motions_HK} 
\begin{center}
\def\arraystretch{1.3}
\setlength{\tabcolsep}{1.6pt}    
    
\small

\begin{tabular}{cccccccc}
 &  &  &  &  &  &  & \tabularnewline
\hline 
\hline 
$H-K_s$ & \#stars & $\mu_l$ & $d\mu_l$ & $\sigma\mu_l$ & $\mu_b$ & $d\mu_b$ & $\sigma\mu_b$\tabularnewline
(mag) &  & (mas/yr) & (mas/yr) & (mas/yr) & (mas/yr) & (mas/yr) & (mas/yr)\tabularnewline
\hline 
1.3-1.5 & 131 & 2.37 & 0.30 & 2.79 & 0.04 & 0.23 & 2.15\tabularnewline
1.5-1.7 & 267 & 1.69 & 0.22 & 2.82 & 0.07 & 0.18 & 2.37\tabularnewline
1.7-1.9 & 439 & 0.53 & 0.16 & 2.67 & 0.04 & 0.14 & 2.42\tabularnewline
1.9-2.1 & 489 & 0.08 & 0.17 & 2.94 & -0.08 & 0.15 & 2.57\tabularnewline
2.1-2.3 & 364 & 0.01 & 0.19 & 2.88 & -0.28 & 0.16 & 2.46\tabularnewline
2.3-2.5 & 247 & -0.12 & 0.21 & 2.63 & 0.08 & 0.20 & 2.46\tabularnewline
2.5-2.7 & 135 & -0.03 & 0.28 & 2.64 & -0.26 & 0.29 & 2.66\tabularnewline
2.7-2.9 & 63 & -0.51 & 0.46 & 2.90 & -0.18 & 0.38 & 2.40\tabularnewline
\hline 

\end{tabular}    
 
\end{center}
\footnotesize
\textbf{Notes.} $H-K_s$, $\mu_i$, $d\mu_i$, and $\sigma\mu_i$ indicate the approximate colour range (see Fig.\,\ref{CMD_pm}) and the median value of the proper motion distribution with its corresponding standard error and the standard deviation, respectively. The subindex $i$ indicates Galactic longitude ($i=l$) or latitude ($i=b$).

 \end{table}

\section{Line-of-sight velocity dispersion}

\begin{table}

\caption{Line-of-sight velocity dispersion for the colour cuts in the CMDs.}
\label{table_vr} 
\begin{center}
\def\arraystretch{1.3}
\setlength{\tabcolsep}{3pt}    
    
\small

\begin{tabular}{cccc|cccc}
 &  &  & \multicolumn{1}{c}{} &  &  &  & \tabularnewline
\hline 
\hline 
$J-K_s$ & \#stars & $\sigma_{vr}$ & $d\sigma_{vr}$ & $H-K_s$ & \#stars & $\sigma_{vr}$ & $d\sigma_{vr}$\tabularnewline
(mag) &  & (mas/yr) & (mas/yr) & (mag) &  & (mas/yr) & (mas/yr)\tabularnewline
\hline 
4-4.5 & 22 & 1.74 & 0.30 & 1.3-1.5 & 21 & 1.93 & 0.35\tabularnewline
4.5-5 & 38 & 2.25 & 0.34 & 1.5-1.7 & 43 & 2.06 & 0.31\tabularnewline
5-5.5 & 57 & 2.26 & 0.29 & 1.7-1.9 & 65 & 2.09 & 0.22\tabularnewline
5.5-6 & 185 & 2.25 & 0.14 & 1.9-2.1 & 222 & 2.16 & 0.11\tabularnewline
6-6.5 & 288 & 2.36 & 0.10 & 2.1-2.3 & 303 & 2.50 & 0.11\tabularnewline
6.5-7 & 226 & 2.53 & 0.10 & 2.3-2.5 & 215 & 2.47 & 0.11\tabularnewline
7-7.5 & 145 & 2.24 & 0.13 & 2.5-2.7 & 134 & 2.15 & 0.12\tabularnewline
- & - & - & - & 2.7-2.9 & 60 & 2.25 & 0.23\tabularnewline
\hline 

\end{tabular}

\end{center}
\footnotesize
\textbf{Notes.} $J-K_s$ and $H-K_s$ indicate the approximate colour range (see Fig.\,\ref{CMD_pm}). $\sigma_{vr}$ and $d\sigma_{vr}$ correspond to the line of sight velocity dispersion and its standard deviation.

 \end{table}

\section{Metallicity analysis}

\begin{table*}

\caption{Results from the GMM analysis of the metallicity distribution using the CMD $K_s$ versus $J-K_s$.}
\label{met_table_JK} 
\begin{center}
\def\arraystretch{1.3}
\setlength{\tabcolsep}{5pt}    
    
\small

\begin{tabular}{ccccccccc}
 &  &  &  &  &  &  & & \tabularnewline
\hline 
\hline 
$J-K_s$ & \#stars & $W_1$ &  $[M/H]_1$ & $\sigma_{[M/H]_1}$ &$W_2$ & $[M/H]_2$ & $\sigma_{[M/H]_2}$ & $W_1/W_2$ \tabularnewline
(mag) &  & (norm. units) & (dex) & (dex) & (norm. units) & (dex) & (dex) & \tabularnewline
\hline 

4-4.5 & 20 & 0.30$\pm$0.15 & -0.44$\pm$0.36 & 0.36$\pm$0.34 & 0.70$\pm$0.15 & 0.24$\pm$0.08 & 0.25$\pm$0.08 & 0.42$\pm$0.23\tabularnewline
4.5-5 & 32 & 0.40$\pm$0.22 & -0.38$\pm$0.52 & 0.36$\pm$0.52 & 0.60$\pm$0.22 & 0.18$\pm$0.12 & 0.31$\pm$0.12 & 0.66$\pm$0.43\tabularnewline
5-5.5 & 49 & 0.42$\pm$0.07 & -0.34$\pm$0.09 & 0.47$\pm$0.09 & 0.58$\pm$0.07 & 0.24$\pm$0.07 & 0.33$\pm$0.07 & 0.72$\pm$0.14\tabularnewline
5.5-6 & 135 & 0.41$\pm$0.04 & -0.19$\pm$0.06 & 0.45$\pm$0.06 & 0.59$\pm$0.04 & 0.27$\pm$0.04 & 0.30$\pm$0.04 & 0.71$\pm$0.09\tabularnewline
6-6.5 & 202 & 0.41$\pm$0.03 & -0.14$\pm$0.05 & 0.45$\pm$0.05 & 0.59$\pm$0.03 & 0.30$\pm$0.03 & 0.29$\pm$0.03 & 0.69$\pm$0.07\tabularnewline
6.5-7 & 143 & 0.38$\pm$0.04 & -0.21$\pm$0.08 & 0.54$\pm$0.08 & 0.62$\pm$0.04 & 0.26$\pm$0.04 & 0.30$\pm$0.04 & 0.60$\pm$0.08\tabularnewline
7-7.5 & 83 & 0.39$\pm$0.07 & -0.03$\pm$0.09 & 0.46$\pm$0.09 & 0.61$\pm$0.07 & 0.37$\pm$0.05 & 0.28$\pm$0.05 & 0.64$\pm$0.13\tabularnewline

\hline 

\end{tabular}    
 
\end{center}
\footnotesize
\textbf{Notes.} $W_i$, $[M/H]_i$, and $\sigma_{[M/H]_i}$, indicate the relative weight, the mean metallicity and the standard deviation of each of the components (i=1 or 2, corresponding to the metal poor and rich components, respectively) of the GMM modelling.

 \end{table*}

\begin{table*}

\caption{Results from the GMM analysis of the metallicity distribution using the CMD $K_s$ versus $H-K_s$.}
\label{met_table_HK} 
\begin{center}
\def\arraystretch{1.3}
\setlength{\tabcolsep}{5pt}    
    
\small

\begin{tabular}{ccccccccc}
 &  &  &  &  &  &  & & \tabularnewline
\hline 
\hline 
$H-K_s$ & \#stars & $W_1$ &  $[M/H]_1$ & $\sigma_{[M/H]_1}$ &$W_2$ & $[M/H]_2$ & $\sigma_{[M/H]_2}$ & $W_1/W_2$ \tabularnewline
(mag) &  & (norm. units) & (dex) & (dex) & (norm. units) & (dex) & (dex) & \tabularnewline
\hline

1.3-1.5 & 19 & 0.36$\pm$0.16 & -0.39$\pm$0.31 & 0.38$\pm$0.31 & 0.64$\pm$0.16 & 0.23$\pm$0.12 & 0.28$\pm$0.12 & 0.56$\pm$0.29\tabularnewline
1.5-1.7 & 36 & 0.35$\pm$0.17 & -0.35$\pm$0.34& 0.43$\pm$0.34 & 0.65$\pm$0.17 & 0.19$\pm$0.10 & 0.29$\pm$0.10 & 0.54$\pm$0.29\tabularnewline
1.7-1.9 & 57 & 0.45$\pm$0.06 & -0.21$\pm$0.08 & 0.44$\pm$0.08 & 0.55$\pm$0.06 & 0.23$\pm$0.07 & 0.32$\pm$0.07 & 0.81$\pm$0.15\tabularnewline
1.9-2.1 & 167 & 0.41$\pm$0.04 & -0.23$\pm$0.06 & 0.47$\pm$0.06 & 0.59$\pm$0.04 & 0.26$\pm$0.04 & 0.31$\pm$0.04 & 0.69$\pm$0.08\tabularnewline
2.1-2.3 & 202 & 0.38$\pm$0.04 & -0.22$\pm$0.05 & 0.49$\pm$0.05 & 0.62$\pm$0.04 & 0.29$\pm$0.03 & 0.30$\pm$0.03 & 0.62$\pm$0.07\tabularnewline
2.3-2.5 & 133 & 0.39$\pm$0.06 & -0.05$\pm$0.08 & 0.52$\pm$0.08 & 0.61$\pm$0.06 & 0.29$\pm$0.05 & 0.29$\pm$0.05 & 0.63$\pm$0.11\tabularnewline
2.5-2.7 & 73 & 0.40$\pm$0.06 & -0.03$\pm$0.08 & 0.41$\pm$0.08 & 0.60$\pm$0.06 & 0.39$\pm$0.05 & 0.27$\pm$0.05 & 0.68$\pm$0.13\tabularnewline
2.7-2.9 & 34 & 0.42$\pm$0.10 & -0.05$\pm$0.11 & 0.37$\pm$0.11 & 0.58$\pm$0.10 & 0.38$\pm$0.08 & 0.25$\pm$0.08 & 0.71$\pm$0.22\tabularnewline
\hline 

\end{tabular}    
 
\end{center}
\footnotesize
\textbf{Notes.} $W_i$, $[M/H]_i$, and $\sigma_{[M/H]_i}$, indicate the relative weight, the mean metallicity and the standard deviation of each of the components (i=1 or 2, corresponding to the metal poor and rich components, respectively) of the GMM modelling.

 \end{table*}

\begin{table*}

\caption{Spatial distribution of stars with different metallicities.}
\label{table_spatial} 
\begin{center}
\def\arraystretch{1.3}
\setlength{\tabcolsep}{5pt}    
    
\small

\begin{tabular}{cccccc|cccccc}
 &  &  &  &  & \multicolumn{1}{c}{} &  &  &  &  &  & \tabularnewline
\hline 
\hline 
$J-K_s$ & \#stars & $\bar{x}$ & $\sigma_{\bar{x}}$ & $\bar{y}$ & $\sigma_{\bar{y}}$ & $H-K_s$ & \#stars & $\bar{x}$ & $\sigma_{\bar{x}}$ & $\bar{y}$ & $\sigma_{\bar{y}}$\tabularnewline
(mag) &  & (arcsec) & (arcsec) & (arcsec) & (arcsec) & (mag) &  & (arcsec) & (arcsec) & (arcsec) & (arcsec)\tabularnewline
\hline 
4-4.5 & 20 & $43\pm33$ & $78\pm8$ & $8\pm14$ & $30\pm3$ & 1.3-1.5 & 19 & $-45\pm28$ & $71\pm8$ & $-7\pm12$ & $31\pm4$\tabularnewline
4.5-5 & 32 & $-24\pm9$ & $56\pm7$ & $-6\pm4$ & $25\pm3$ & 1.5-1.7 & 36 & $-14\pm10$ & $61\pm6$ & $-2\pm5$ & $24\pm3$\tabularnewline
5-5.5 & 49 & $-2\pm8$ & $56\pm4$ & $-2\pm2$ & $26\pm2$ & 1.7-1.9 & 57 & $-9\pm6$ & $56\pm4$ & $2\pm4$ & $27\pm2$\tabularnewline
5.5-6 & 135 & $-10\pm1$ & $41\pm3$  & $-2\pm1$ & $21\pm2$ & 1.9-2.1 & 167 & $-10\pm1$ & $40\pm3$ & $-2\pm2$ & $18\pm1$\tabularnewline
6-6.5 & 202 & $-7\pm2$ & $36\pm2$ & $-1\pm1$ & $20\pm1$ & 2.1-2.3 & 202 & $-5\pm2$ & $35\pm2$ & $-3\pm1$ & $20\pm1$\tabularnewline
6.5-7 & 143 & $-2\pm3$ & $41\pm3$ & $0\pm3$ & $20\pm1$ & 2.3-2.5 & 133 & $5\pm2$ & $43\pm3$ & $3\pm2$ & $22\pm1$\tabularnewline
7-7.5 & 83 & $4\pm3$ & $46\pm4$ & $8\pm3$ & $21\pm2$ & 2.5-2.7 & 73 & $1\pm5$ & $42\pm4$ & $7\pm3$ & $22\pm2$\tabularnewline
- & - &  & - &  &  & 2.7-2.9 & 34 & $-4\pm6$ & $35\pm6$ & $-2\pm7$ & $18\pm1$\tabularnewline
\hline 

\end{tabular}

\end{center}
\footnotesize
\textbf{Notes.} $\bar{x}$, $\sigma_{\bar{x}}$, $\bar{y}$, and $\sigma_{\bar{y}}$, indicate the median position and the standard deviation of the stars within each colour cut for the $x$ and $y$ coordinates. 
 \end{table*}

\end{document}